\documentclass[letterpaper]{aa530}
\usepackage{txfonts}
\usepackage{graphicx}
\usepackage{natbib}
\bibpunct{(}{)}{;}{a}{}{,}

\newcommand{\teff}   {T$_{\rm eff}$}
\newcommand{\vsini}  {{\it v}\,sin\,{\it i}}
\newcommand{\vrad}   {{v$_r$}}
\newcommand{\kms}    {km\,s$^{-1}$}
\newcommand{\ms}     {m\,s$^{-2}$}
\newcommand{\deltav} {$\Delta v$}

\defcitealias{mora2002}{Paper~I}

\begin{document}
\title{Dynamics of the circumstellar gas in the Herbig Ae stars
       BF~Orionis, SV~Cephei, WW~Vulpeculae and XY~Persei}

\author{A.~Mora         \inst{1}        \and
        C.~Eiroa        \inst{1,2}      \and
        A.~Natta        \inst{3}        \and
        C.A.~Grady      \inst{4}        \and
        D.~de Winter    \inst{5}        \and
        J.K.~Davies     \inst{6}        \and
        R.~Ferlet       \inst{7}        \and
        A.W.~Harris     \inst{8}        \and
        L.F.~Miranda    \inst{9}        \and
        B.~Montesinos   \inst{10,9}     \and
        R.D.~Oudmaijer  \inst{11}       \and
        J.~Palacios     \inst{1}        \and
        A.~Quirrenbach  \inst{12}       \and
        H.~Rauer        \inst{8}        \and
        A.~Alberdi      \inst{9}        \and
        A.~Cameron      \inst{13}       \and
        H.J.~Deeg       \inst{14}       \and
        F.~Garz\'on     \inst{14}       \and
        K.~Horne        \inst{13}       \and
        B.~Mer\'{\i}n   \inst{10}       \and
        A.~Penny        \inst{15}       \and
        J.~Schneider    \inst{16}       \and
        E.~Solano       \inst{10}       \and
        Y.~Tsapras      \inst{13}       \and
        P.R.~Wesselius  \inst{17}}

\institute{
Departamento de F\'{\i}sica Te\'orica C-XI, Universidad Aut\'onoma de Madrid, Cantoblanco 28049 Madrid, Spain
        \and
Visiting Scientist at ESA/ESTEC and Leiden University, The Netherlands
        \and
Osservatorio Astrofisico di Arcetri, Largo Fermi 5, I-50125 Firenze, Italy
        \and
NOAO/STIS, Goddard Space Flight Center, Code 681, NASA/GSFC, Greenbelt, MD 20771, USA
        \and
TNO/TPD-Space Instrumentation, Stieltjesweg 1, PO Box 155, 2600 AD Delft, The Netherlands
        \and
Astronomy Technology Centre, Royal Observatory, Blackford Hill, Edinburgh, UK
        \and
CNRS, Institute d'Astrophysique de Paris, 98bis Bd. Arago, 75014 Paris, France 
        \and
DLR Department of Planetary Exploration, Rutherfordstrasse 2, 12489 Berlin, Germany
        \and
Instituto de Astrof\'{\i}sica de Andaluc\'{\i}a, Apartado de Correos 3004, 18080 Granada, Spain
        \and
LAEFF, VILSPA, Apartado de Correos 50727, 28080 Madrid, Spain
        \and
Department of Physics and Astronomy, University of Leeds, Leeds LS2 9JT, UK
        \and
Department of Physics, Center for Astrophysics and Space Sciences, University of California San Diego, Mail Code 0424, La Jolla, CA 92093-0424, USA
        \and
Physics \& Astronomy, University of St. Andrews, North Haugh, St. Andrews KY16 9SS, Scotland, UK
        \and
Instituto de Astrof\'{\i}sica de Canarias, La Laguna 38200 Tenerife, Spain
        \and
Rutherford Appleton Laboratory, Didcot, Oxfordshire OX11 0QX, UK
        \and
Observatoire de Paris, 92195 Meudon, France
        \and
SRON, Universiteitscomplex ``Zernike'', Landleven 12, P.O. Box 800, 9700 AV Groningen, The Netherlands
}

\offprints{Alcione Mora, \\
\email{alcione.mora@uam.es}}

\date{Received 3 June 2003 / Accepted 24 February 2004}


\abstract{
We present high resolution ($\lambda$/$\Delta \lambda$~=~49\,000) \'echelle spectra of the intermediate mass, pre-main sequence stars \object{BF Ori}, \object{SV Cep}, \object{WW Wul} and \object{XY Per}.
The spectra cover the range 3800~--~5900~\AA\ and monitor the stars on time scales of months and days.
All spectra show a large number of Balmer and metallic lines with variable blueshifted and redshifted absorption features superimposed to the photospheric stellar spectra.
Synthetic Kurucz models are used to estimate rotational velocities, effective temperatures and gravities of the stars.
The best photospheric models are subtracted from each observed spectrum to determine the variable absorption features due to the circumstellar gas; those features are characterized in terms of their velocity, $v$, dispersion velocity, $\Delta v$, and residual absorption, $R_{\rm max}$.
The absorption components detected in each spectrum can be grouped by their similar radial velocities and are interpreted as the signature of the dynamical evolution of gaseous clumps with, in most cases, solar-like chemical composition.
This infalling and outflowing gas has similar properties to the circumstellar gas observed in UX~Ori, emphasizing the need for detailed theoretical models, probably in the framework of the magnetospheric accretion scenario, to understand the complex environment in Herbig Ae (HAe) stars.
WW Vul is unusual because, in addition to infalling and outflowing gas with properties similar to those observed in the other stars, it shows also transient absorption features in metallic lines with no obvious counterparts in the hydrogen lines.
This could, in principle, suggest the presence of CS gas clouds with enhanced metallicity around WW~Vul.
The existence of such a metal-rich gas component, however, needs to be confirmed by further observations and a more quantitative analysis.

}

\titlerunning{Dynamics of the circumstellar gas in the HAe stars BF~Orionis, SV~Cephei, WW~Vulpeculae and XY~Persei}
\authorrunning{Mora et al.}

\maketitle


\section{Introduction}

Observations reveal that the dynamics of the circumstellar (CS) gaseous disks around intermediate mass, young main sequence (MS) and pre-main sequence (PMS) stars is extremely complex.
Variable absorption components detected in many lines of different elements and ions constitute good examples of such complexity.
The kinematics and intensity strength of the absorption components contain relevant information on the physical properties of the gas.
Further, their detailed characterization and analysis provide clues and constraints on plausible formation mechanisms as well as on theoretical scenarios describing the structure and nature of the CS gaseous disks.

The presence of metal-rich planetesimals in the young MS  $\beta$~Pic system has been inferred both observationally and theoretically in a long series of papers \citep[e.g.][and references therein]{lagrange2000}.
Summarizing, there are two main arguments on which this inference is based.
Firstly, dust causing the far-IR excess and also seen in the $\beta$~Pic disk images may be second generation material continuously replenished by collisions of large solid bodies or slow evaporation. 
Secondly, transient spectral line absorptions, usually redshifted, of different chemical species in a wide range of ionization states can be modelled in terms of the evaporation of km-sized bodies on star-grazing orbits.
Star-grazing planetesimals have also been suggested to exist in the 51~Oph system, a star with an uncertain evolutionary status \citep{roberge2002,vandenancker2001}; also, blueshifted absorption in excited fine structure lines of \ion{C}{ii}$^*$ at 1037~\AA\ and \ion{N}{ii}$^*$ at 1085 and 1086~\AA\ have been used to infer the presence of $\sim$1~m bodies in the Vega-type binary MS system $\sigma$~Her \citep{chen2003}.  

Absorption features similar to those observed in $\beta$~Pic have been observed towards many HAe stars (see Natta et al. 2000 for the description of these stars as a subgroup of HAeBe stars), particularly in the UXOR-subclass \citep[e.g.][]{grady1996} and, by analogy, they have been interpreted as indicative of the presence of large solid bodies in the CS disks  around these PMS stars \citep[e.g.][and references therein]{grady2000}.
In principle, this interpretation is not in conflic with the accepted time scale for the formation of planetesimals of $\sim$10$^4$~yr \citep{beckwith2000}, which suggests that planetesimals should already be present during the PMS phase of stars ($\sim$1-10~Myr).
This explanation for the variable absorption features observed in HAe stars is, however, controversial and, in fact, \citet{grinin1994} pointed out other alternatives, more concretely dissipation of dust clouds and the simultaneous infall of cool gas onto the star.
\citet{natta2000} have analyzed the chemical composition of a strong redshifted event in UX~Ori and shown it to have a solar-like composition.
Instead of the planetesimal origin for the transient components, those authors suggested gas accretion from a CS disk.
This result is supported by \citet[from now on Paper~I]{mora2002} who presented the analysis of a large series of high resolution optical spectra of UX~Ori.
Many variable absorption events in hydrogen and metallic lines were attributed to the dynamical evolution of gaseous clumps with non metal-rich, roughly solar chemical compositions.
In addition, \citet{beust2001} have shown that the  $\beta$~Pic infalling planetesimal model would not produce detectable absorptions in typical PMS HAe CS conditions.
We also note that dust disks around HAe stars are primordial and can be explained in the context of irradiated PMS CS disk models \citep{natta2001}.
 
In this paper we present high resolution spectra of the HAe stars BF~Ori, SV~Cep, WW~Vul and XY~Per and perform an analysis similar to that carried out for UX~Ori \citepalias{mora2002}.
The spectra show very active and complex CS gas in these objects; many transient absorption features in hydrogen and metallic lines are detected, indicating similar properties of the gas around these stars to those of UX~Ori CS gas.
In addition, the spectra of WW Vul show metallic features without obvious hydrogen counterparts; in this sense, this star presents a peculiar behaviour.
The layout of the paper is as follows: Section~\ref{observations} presents a brief description of the observations.
Section~\ref{analysis} presents the results and an analysis of the photospheric spectra and the CS contribution.
Section~\ref{discussion} presents a short discussion on the kinematics and strength of the variable features, and on the metallic events detected in WW~Vul.
Finally, Sect.~\ref{conclussions} gives some concluding remarks.

\section{Observations}
\label{observations}

High resolution \'echelle spectra were collected with the Utrecht Echelle Spectrograph (UES) at the 4.2~m WHT (La~Palma observatory) during four observing runs in May, July and October 1998 and January 1999. 
28 spectra were obtained: 4 of BF~Ori, 7 of SV~Cep, 8 of WW~Vul and 9 of XY~Per (XY~Per is a visual binary sistem with a Herbig Ae primary and a B6Ve secondary separated by 1.4'', the good seeing during the observations allowed us to fully separate both components, in this paper only the HAe primary star has been studied). 
The wavelength range was 3800-5900~{\AA} and the spectral resolution, $\lambda / \Delta\lambda$,~=~49\,000 (6~\kms).
Wavelength calibration was performed using Th-Ar arc lamp spectra.
Typical errors of the wavelength calibration are $\sim$5 times smaller than the spectral resolution.
The observing log, exposure times and signal to noise ratio (SNR) values, measured at $\lambda \simeq 4680$~\AA, are given in Table~\ref{observing_log}.
Further details of the observations and reduction procedure are given by \citet{mora2001}.
For some spectra there are simultaneous optical photo-polarimetric and near-IR photometric observations \citep{oudmaijer2001,eiroa2001}.
Table~\ref{observing_log} also presents these simultaneous data.
At the time of the observations the stars were close to their brightest state, BF~Ori and XY~Per, or at average brightness, SV~Cep and WW~Vul \citep{herbst1999,eiroa2002}.

\begin{table}
\caption{EXPORT UES/WHT observing logs of BF~Ori, SV~Cep, WW~Vul and XY~Per.
The Julian date ($-$2\,450\,000) of each spectrum is given in Col. 1.
Column 2 shows the exposure time in seconds.
Column 3 gives the SNR at $\lambda \simeq 4680$~\AA.
Columns 4 to 6 give simultaneous $V$, \%P$_V$ and $K$ photopolarimetric data, where available.
Typical errors are 0.10 in $V$, 0.05 in $K$ and 0.1\% in \%P$_V$.}
\label{observing_log}
\centerline{
\begin{tabular}{lrrlll}
\hline
\hline
{\bf BF Ori} &                 &     &        &         &      \\
\hline
Julian date  &t$_{\rm exp}$ (s)& SNR & $V$    & \%P$_V$ & $K$  \\
\hline
1112.6324    & 1800            & 280 &  9.65  &  0.56   & 7.91 \\
1113.6515    & 2700            & 190 &  9.79  &  0.75   & 7.85 \\
\hline
1209.5542    & 2700            & 210 &  --    &  --     & 7.76 \\
1210.4571    & 2700            & 280 &  --    &  0.14   & --   \\
\hline
             &                 &     &        &         &      \\
\hline
\hline
{\bf SV Cep} &                 &     &        &         &      \\
\hline
Julian date  &t$_{\rm exp}$ (s)& SNR & $V$    & \%P$_V$ & $K$  \\
\hline
950.6668     & 1800            & 100 & --     & --      & --   \\
950.6893     & 1800            & 120 & --     & --      & --   \\
\hline
1025.6260    & 2700            & 140 & --     & 0.96    & --   \\
1025.6595    & 1800            & 130 & --     & --      & --   \\
1026.6684    & 2700            & 140 & --     & --      & --   \\
\hline
1113.4730    & 2700            & 70  & 11.01  & 1.05    & --   \\
\hline
1209.3372    & 2700            & 170 & --     & --      & --   \\
\hline
             &                 &     &        &         &      \\
\hline
\hline
{\bf WW Vul} &                 &     &        &         &      \\
\hline
Julian date  &t$_{\rm exp}$ (s)& SNR & $V$    & \%P$_V$ & $K$  \\
\hline
950.6176     & 1800            & 130 &  --    &  --     & --   \\
950.6413     & 1800            & 130 &  10.89 &  0.69   & --   \\
951.6232     & 1800            & 120 &  --    &  --     & --   \\
951.6465     & 1800            & 150 &  --    &  --     & --   \\
\hline
1023.5186    & 1800            & 140 &  --    &  --     & 7.37 \\
1023.5423    & 2700            & 190 &  --    &  0.40   & --   \\
\hline
1112.3689    & 1800            & 110 &  10.77 &  0.37   & 7.44 \\
1113.3958    & 2700            & 120 &  11.03 &  0.65   & 7.50 \\
\hline
             &                 &     &        &         &      \\
\hline
\hline
{\bf XY Per} &                 &     &        &         &      \\
\hline
Julian date  &t$_{\rm exp}$ (s)& SNR & $V$    & \%P$_V$ & $K$  \\
\hline
1024.6728    & 1800            & 200 & --     & --      & --   \\
1024.6967    & 1800            & 170 & 9.04   & 1.49    & --   \\
1025.6948    & 1800            & 230 & --     & 1.55    & --   \\
1025.7171    & 1800            & 270 & --     & --      & --   \\
1026.7065    & 1800            & 220 & --     & --      & --   \\
\hline
1112.4978    & 1800            & 270 & 9.12   & 1.65    & 5.97 \\
1113.5483    & 2700            & 230 & 9.05   & 1.53    & 5.99 \\
\hline
1207.3786    & 2700            & 190 & --     & --      & --   \\
1209.3844    & 1800            & 260 & 9.51   & 1.58    & 6.18 \\
\hline
\end{tabular}}
\end{table}

\section{Analysis of the spectra and results}
\label{analysis}

\subsection{The photospheric spectra}
\label{the_photospheric_spectra}

Circumstellar absorptions with complex profiles and blended components are detected in hydrogen and metallic lines in all UES spectra of the four stars. 
The analysis of those spectral features requires the subtraction of the underlying photospheric spectra.
Such subtraction is carried out following the method outlined in detail in \citetalias{mora2002}.
Briefly, \citet{kuruczCD13} model atmospheres are used to synthesize photospheric spectra.
Four parameters are estimated: the heliocentric radial velocity (\vrad), the rotation velocity (\vsini), the effective temperature (\teff) and the surface gravity (log~g).
It has been shown by \citet{grinin2001} that the photospheric lines of BF Ori and WW Vul can be well reproduced using solar metallicity synthetic spectra.
On the other hand, the presence of CS components in most of the lines makes it very difficult to carry out a detailed abundance analysis.
We have thus decided to assume that the four stars have solar metallicities.
The atomic line data have been obtained from the VALD database \citep{kupka1999}.
The `best' synthetic model, defined by the parameters listed before, is selected by comparing some appropriate faint photospheric absorption lines among the observed spectra and the synthetic ones.
This is not straightforward because each star behaves differently, and the choice of pure photospheric lines in the spectra of such highly variable objects is not trivial.
Therefore, slightly different, ad-hoc approaches for each star are needed.
These approaches are discussed below and the stellar parameters giving the best synthetic spectra are given in Table~\ref{parameters}.
They are compatible within the uncertainties with the spectral types and rotational velocities quoted by \citet{mora2001} and with the results by \citet{grinin2001}, who studied BF~Ori and WW~Vul.   

\begin{table}
\caption{Stellar parameters defining the `best'  synthetic Kurucz models for each star.
See Sect.~\ref{the_photospheric_spectra} for a discussion on the uncertainties of \teff, log~g and \vsini.}
\label{stellar_parameters}
\centerline{
\begin{tabular}{lcccc}
\hline
\hline
Star   & 
\parbox{0.8cm}{\centering {\teff} \\ (K)} & 
log~g & 
\parbox{1cm}{\centering \vsini\ \\ (\kms)} & 
\parbox{1cm}{\centering \vrad\  \\ (\kms)} \\
\hline
BF~Ori & 8750  & 3.5 & 37  &  23.1 $\pm$ 1.9 \\
SV~Cep & 10000 & 4.0 & 225 & -11.9 $\pm$ 0.8 \\
WW~Vul & 9000  & 4.0 & 210 & -10.4 $\pm$ 1.2 \\
XY~Per & 8500  & 3.5 & 200 &   8.3 $\pm$ 0.6 \\
\hline
\end{tabular}}
\label{parameters}
\end{table}

\begin{figure*}
\centerline{
\includegraphics[clip=true,angle=-90,width=\hsize]{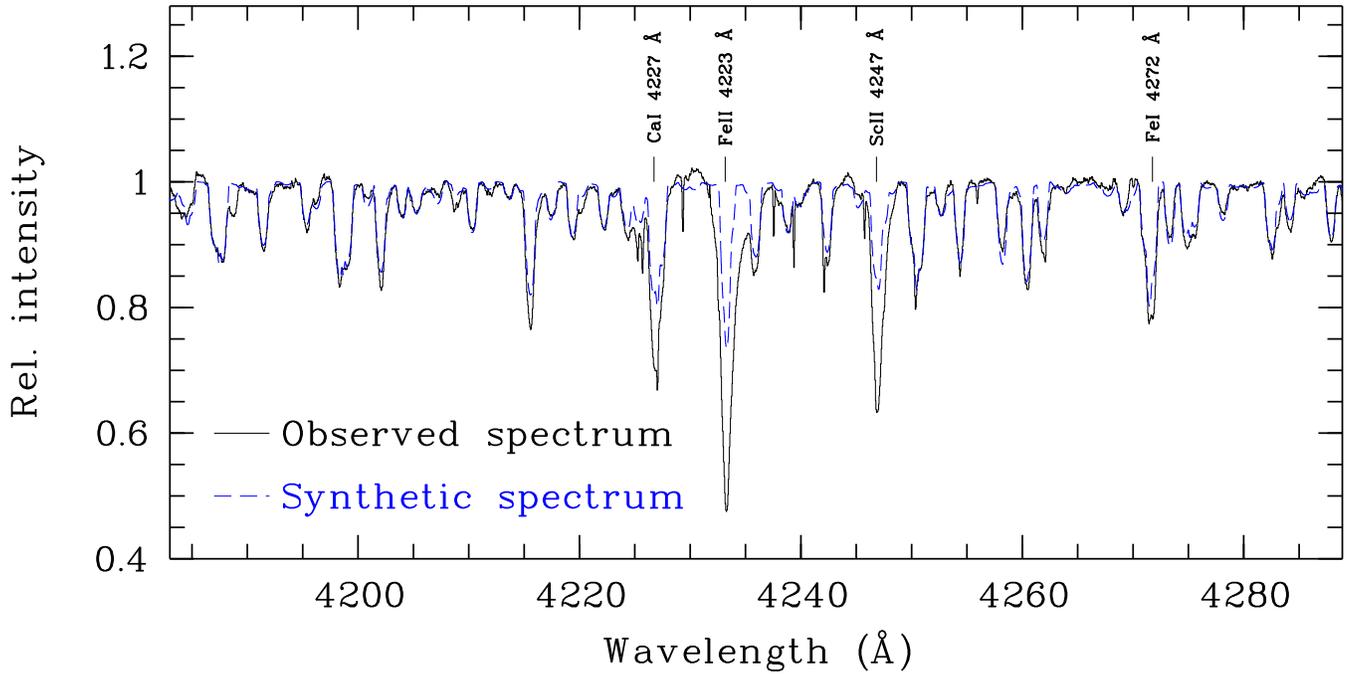}}
\caption{Synthetic (dashed line) vs observed spectrum (solid line) for BF~Ori.
The agreement between the photospheric and the observed median spectrum is remarkable over a large wavelength range.
The extra absorption in the stronger lines (identified in the figure) is due to the circumstellar contribution.
This example illustrates the need to use very faint lines to estimate the stellar parameters.
(This figure is available in color in electronic form)}
\label{bfori_obs_vs_syn}
\end{figure*}

{\em BF~Ori:}
The photospheric lines are narrow and do not show a noticeable variability.
\vrad\ is estimated using faint photospheric lines and its rms error is low.
Estimated errors are  250~K  for \teff\ (the step in the synthetic spectrum grid created using Kuruzc's models), and 6~\kms\ for \vsini\ (the spectral resolution); the log~g values considered have been restricted to 3.5 and 4.0.
Fig.~\ref{bfori_obs_vs_syn} shows the excellent agreement between the synthetic  model and the observed median spectrum.
The extra absorption seen in the stronger lines is due to the circumstellar contribution.  

{\em SV~Cep:}
According to \citet{finkenzeller1984} the radial velocity of HAe stars coincides within a few \kms with the radial velocity of interstellar (IS) lines.
The strong \ion{Na}{i}~D IS components in the line of sight of SV~Cep allow us to determine a precise value of the radial velocity of this star.
Its high temperature, rotation velocity and variability make it extremely difficult to identify photospheric lines in order to derive the Kurucz stellar parameters.
We have circumvented this problem by using the EXPORT intermediate resolution spectra \citep{mora2001}, which have very high quality H$\alpha$ profiles and are of great help in making a first estimate of \teff\ and log~g.
>From the H$\alpha$ wings and the almost complete absence of weak photospheric lines in the high resolution spectra, \teff~$\simeq$~10\,000~K and log~g~=~4.0 are estimated.
It was assumed that \vsini~=~206~\kms \citep{mora2001}.
The adequacy of this choice can be seen in Fig.~\ref{svcep_obs_vs_syn}, where the 
observed median SV~Cep H$\alpha$ line and 2 synthetic Balmer profiles (\teff~=~10\,000~K, log~g~=~4.0, \vsini~=~206~\kms\ and \teff~=~10\,000~K, log~g~=~3.5, \vsini~=~206~\kms) are shown.
Using those values we have identified 7 blends of weak photospheric lines (absorption intensities $\leq$~4\% of the continuum) with very low variability.
The blends are at $\sim$~3913, 4129, 4176, 4314, 4534, 5041 and 5056~\AA.
A grid of photospheric spectra has been generated with different values of \teff\ and \vsini\ (log~g was assumed to be 4.0 from the $H\alpha$ analysis) and the rms differences between the synthetic and observed spectra have been estimated.
The lowest rms difference is obtained for \teff~$=$~10\,000~K and \vsini~$=$~225~\kms.
The differences between the H$\alpha$ wing profiles broadened to \vsini~$=$~225~\kms\ and \vsini~$=$~206~\kms\ are negligible, so it was not needed to compute log~g again.
Fig.~\ref{svcep_obs_vs_syn} also shows the best synthetic and the observed median spectra for the selected blends. 
Uncertainties of $\sim$500~K (the step of the Kurucz's models) and $\sim$10\% for \teff\ and \vsini\,, respectively, are estimated. 

\begin{figure*}
\centerline{
\includegraphics[clip=true,angle=-90,width=0.5\hsize]{4058.f2a}
\includegraphics[clip=true,angle=-90,width=0.5\hsize]{4058.f2b}}
\caption{
Synthetic vs observed spectra for SV~Cep.
Left: comparison of the intermediate resolution H$\alpha$ spectrum and two synthetic H$\alpha$ profiles.
The best result is obtained for \teff~=~10\,000~K and log~g~=~4.0 (blue dashed line). 
It was assumed that \vsini~=~206~\kms\ \citep{mora2001}.
Right: Median UES spectra (solid lines) of several blends compared with the best synthetic spectra, with \teff~=~10\,000~K, \vsini~=~225~\kms (blue dashed lines).
(This figure is available in color in electronic form)}
\label{svcep_obs_vs_syn}
\end{figure*}

\begin{figure}
\centerline{
\includegraphics[clip=true,angle=-90,width=\hsize]{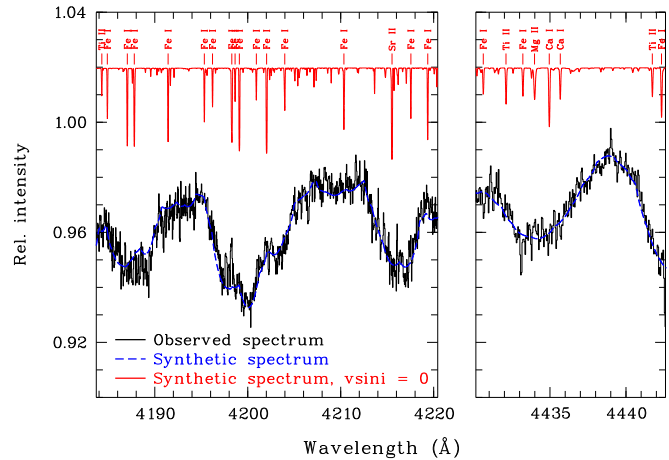}}
\caption{
The observed median UES spectrum of WW~Vul (black continuous line) compared to the broadened synthetic one (blue dashed lines) in two different spectral regions.
At the top of the figure, the unbroadened Kurucz model is shown with some line identifications (red solid line).  
(This figure is available in color in electronic form)}
\label{wwvul_obs_vs_syn}
\end{figure}

{\em WW~Vul:}
\vrad ~is estimated from the sharp \ion{Na}{i}~D IS lines.
The star was very active during the observing runs and its spectra show a large number of broad and variable absorption features, e.g. redshifted features are superimposed on practically each photospheric line in the JD 1113.40 spectrum, which pose severe difficulties for the selection of appropriate photospheric lines.
Nevertheless \vsini\ has been estimated with the \ion{Mg}{ii} 4481~\AA\ blended doublet. 
Many pairs of \teff~--~log~g values can reproduce the observed spectra, though the agreement between the observed and synthetic spectra is generally poor.
We have selected \teff~=~9000~K, log~g~=~4.0 as representative of the WW~Vul photosphere because of its compatibility with the results by \citet{mora2001}, though large errors are likely, but not larger than around 10\% in both \teff\ and \vsini.
Fig.~\ref{wwvul_obs_vs_syn} shows a comparison between the observed median spectrum and the broadened synthetic one in two spectral regions where a good fit is achieved; this figure also shows the unbroadened Kurucz's model with some line identifications.
 
\begin{figure*}
\centerline{
\includegraphics[clip=true,angle=-90,width=\hsize]{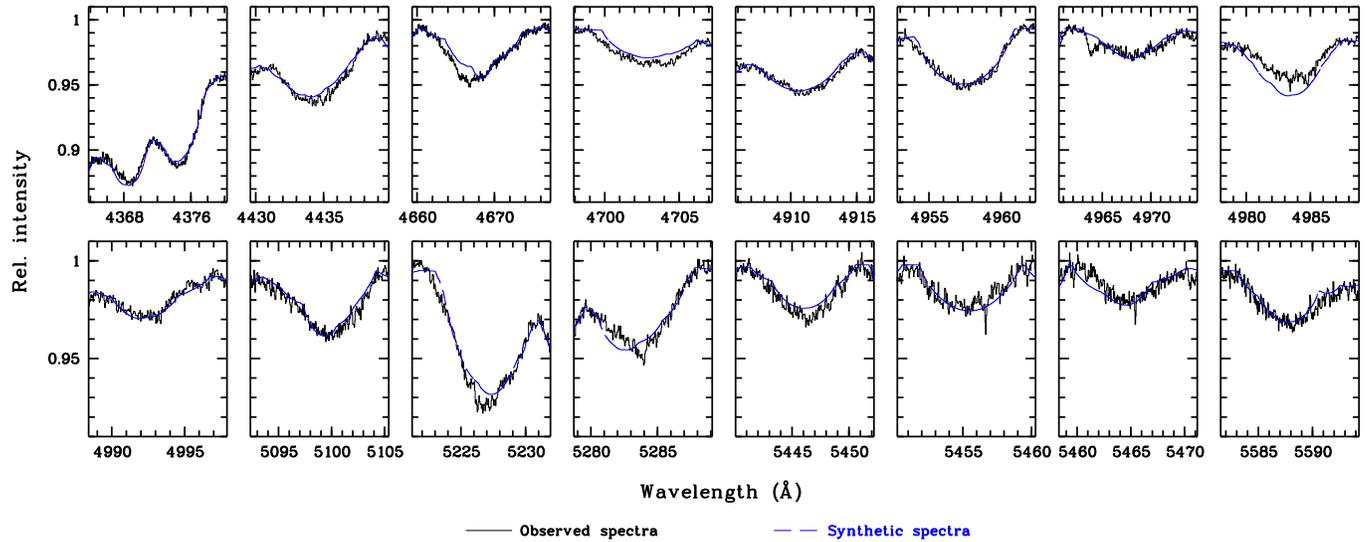}}
\caption{
Synthetic (dashed lines) vs observed median (continuous lines) spectra of XY~Per for 16 spectral features with very low CS contribution.
The stellar parameters of the synthetic spectrum are given in Table~\ref{stellar_parameters}.
(This figure is available in color in electronic form)}
\label{xyper_obs_vs_syn}
\end{figure*}

{\em XY~Per:}
\vrad ~is estimated from the \ion{Na}{i}~D IS components.
The photospheric lines are very broad but the high SNR of the spectra and the relatively low \teff ~allow us to identify a large number of faint line blends in order to perform a precise estimate of \teff, log~g and \vsini ~(16 faint blends with absorption intensities $\leq$~8\% of the continuum with very little CS activity could be identified).    
Errors of \teff\ and log~g are of the order of the step in the Kurucz models, 250~K and 0.5 respectively, while the error in  \vsini\ is very low, $<$10\%.
The comparison between the synthetic and the observed median spectra of the 16 blends is shown in Fig.~\ref{xyper_obs_vs_syn}.

\subsection{The circumstellar transient absorption contribution}
\label{circumstellar_contribution}

Once the best photospheric spectrum of each star is determined, the circumstellar contribution to each observed spectrum can be estimated by subtracting the synthetic one.
The residual spectra show transient absorption features, which can be characterized by means of the normalized residual absorption, defined as \mbox{$R = 1 - F_{\rm obs} / F_{\rm syn}$} \citep{natta2000}.
The $R$ profile of each line reflects the blending of several components.
A multigaussian fit providing the radial velocity, velocity dispersion and absorption strength is used to identify the individual components \citepalias[see][for details]{mora2002}.
Broad redshifted and blueshifted absorptions at different radial velocities are found in the Balmer and metallic lines for all 4 stars analyzed.
We apply the multigaussian fit to Balmer lines (H$\beta$ 4861~\AA, H$\gamma$ 4340~\AA, H$\delta$ 4102~\AA, H$\epsilon$ 3970~\AA, H$\zeta$ 3889~\AA), \ion{Ca}{ii}~K 3934~\AA, \ion{Ca}{ii}~H 3968~\AA, \ion{Na}{i}~D2 5890~{\AA} and \ion{Na}{i}~D1 5896~\AA, as well as to fainter metallic lines \ion{Fe}{ii}~42 multiplet (\mbox{a6S-z6Po} triplet: 4924~\AA, 5018~\AA\ and  5169~\AA), \ion{Ti}{ii} 4444~\AA, \ion{Ti}{ii} 4572~\AA, \ion{Fe}{i} 4046~\AA, \ion{Sc}{ii} 4247~\AA\ and \ion{Ca}{i} 4227~\AA.
We have chosen these ionic lines because they show significant CS variability and are relatively strong and isolated.
Narrow IS components (mainly \ion{Na}{i}, \ion{Ca}{ii} and \ion{Fe}{ii}) with the stellar radial velocity are also detected for all the stars. 

Consecutive spectra with a time delay of $\sim$1~hour of SV~Cep, XY~Per and WW~Vul were taken on several nights (see Table 1).
These spectra were quite similar and  the gaussian deconvolution of the $R$ profiles essentially provides the same values for the fit parameters; thus, any significant variation of the phenomena causing the transient absorptions is excluded on this time scale, at least during these observing periods.
This result gives us confidence in the identification of the components and allows us to add the spectra taken during the same night in order to increase the SNR.
Tables~\ref{master_table_bfori},~\ref{master_table_svcep},~\ref{master_table_wwvul}~and~\ref{master_table_xyper}, available at the CDS, give the radial velocity shift $v$, the velocity dispersion \deltav\ and the absorption strength $R_{\rm max}$, the peak of the $R$ profile, of each identified broad transient absorption component of the lines listed above for BF~Ori, SV~Cep, WW~Vul and XY~Per, respectively.
Column~1 gives the the corresponding Balmer or metallic line, Col.~2 gives the Julian Date, Col.~3 represents the event assigned to the particular absorptions, Col.~4 gives the radial velocity shift $v$, Col.~5 lists the velocity dispersion \deltav\ and Col.~6 gives the absorption strength $R_{\rm max}$.
JD values in Tables~\ref{master_table_svcep},~\ref{master_table_wwvul}~and~\ref{master_table_xyper} correspond to the starting time of the first spectrum of each night. In the following, whenever a JD is given, 2\,450\,000 is subtracted.

\begin{figure*}
\centerline{
\includegraphics[clip=true,angle=-90,width=0.75\hsize]{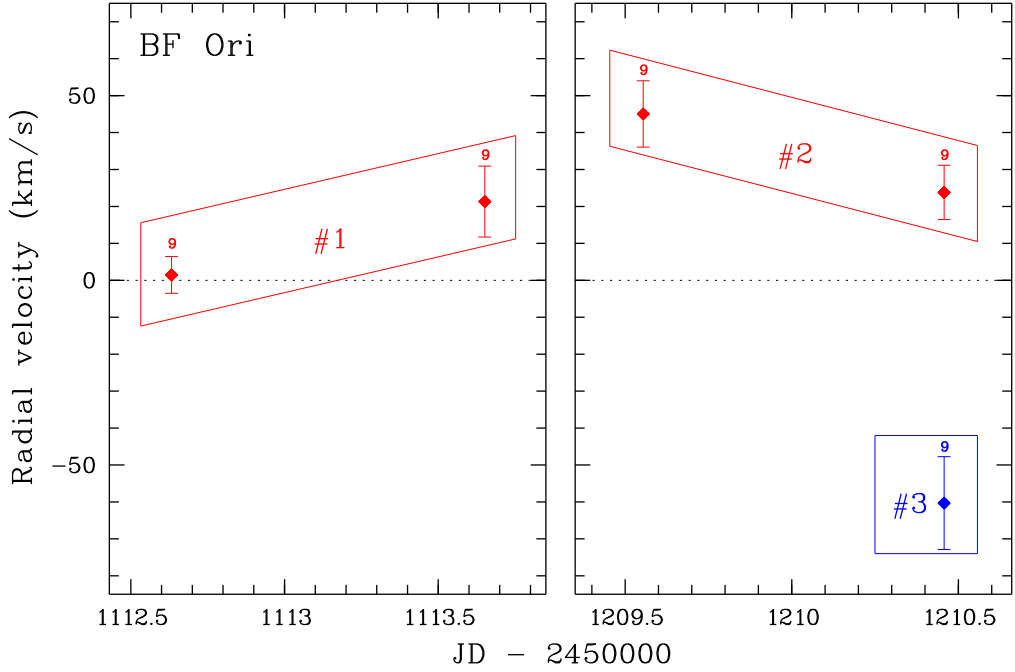}}
\caption{
Events in BF~Ori.
Each point corresponds to the radial velocity of one TAC and represents the average velocity \mbox{\boldmath $<$$v$$>$} of the absorptions with similar radial velocities detected in different lines.
$<$$v$$>$ is a weighted average in which the higher intensity non-blended lines are assigned a weight of 1 and the rest 1/2.
Error bars show the rms error of the average velocity; the numbers above the data points indicate the weighted number of lines used to estimate the average.
Fractional numbers arise from the 1/2 weight attributed to some lines (see text).
Redshifted events (infalling gas) are printed in red colour, while blueshifted events (outflowing gas) are in blue.
(This figure is available in color in electronic form)}
\label{master_bfori}
\end{figure*}

\begin{figure*}
\centerline{
\includegraphics[clip=true,angle=-90,width=0.9\hsize]{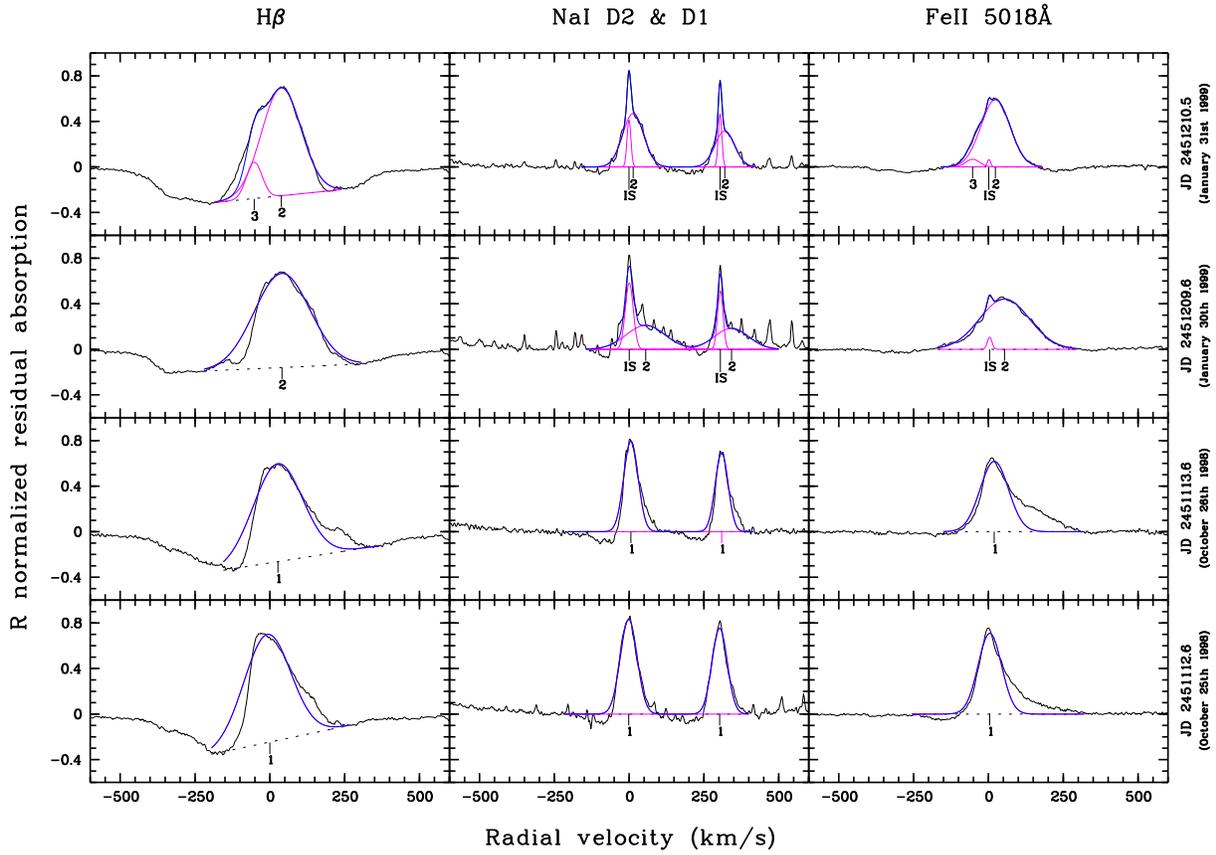}}
\caption{
BF~Ori $R$ profiles.
The normalized residual absorption profiles ($R$ = 1 - $F_{\rm obs}$ / $F_{\rm syn}$) of H$\beta$ (left), \ion{Na}{i} D2 and D1 (middle) and \ion{Fe}{ii}~5018~\AA\ (right) are shown in the figure (black colour).
The corresponding spectra are indicated in the vertical right axis (Julian and civilian epochs).
The identified TACs (gaussian components, pink colour) and the reconstructed $R$ profile fit (blue colour) are displayed.
Event numbers are  shown under the gaussians.
The zero velocity interstellar components are marked as ``IS''.
(This figure is available in color in electronic form)}
\label{bfori_spectra}
\end{figure*}

\begin{figure*}
\centerline{
\includegraphics[clip=true,angle=-90,width=0.75\hsize]{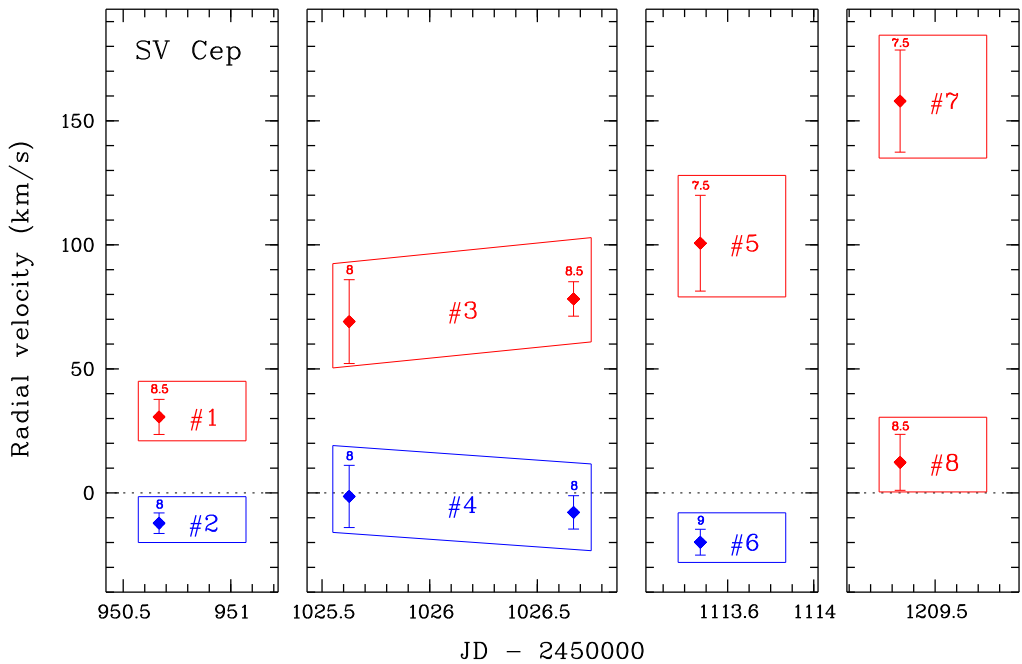}}
\caption{
Events in SV~Cep.
Details as for Fig.~\ref{master_bfori}
(This figure is available in color in electronic form)}
\label{master_svcep}
\end{figure*}

\begin{figure*}
\centerline{
\includegraphics[clip=true,angle=-90,width=0.9\hsize]{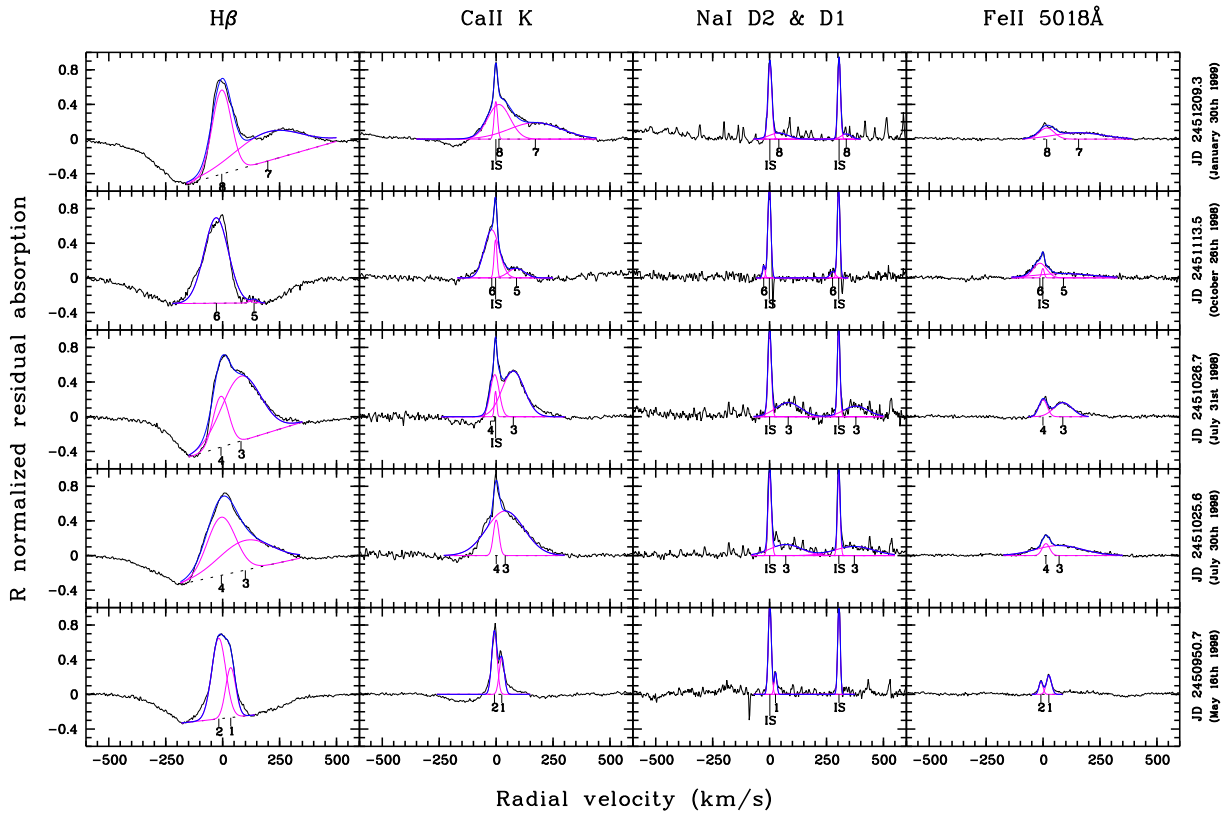}}
\caption{
SV~Cep $R$ profiles.
Details as for Fig~\ref{bfori_spectra}.
(This figure is available in color in electronic form)}
\label{svcep_spectra}
\end{figure*}

\begin{figure*}
\centerline{
\includegraphics[clip=true,angle=-90,width=0.75\hsize]{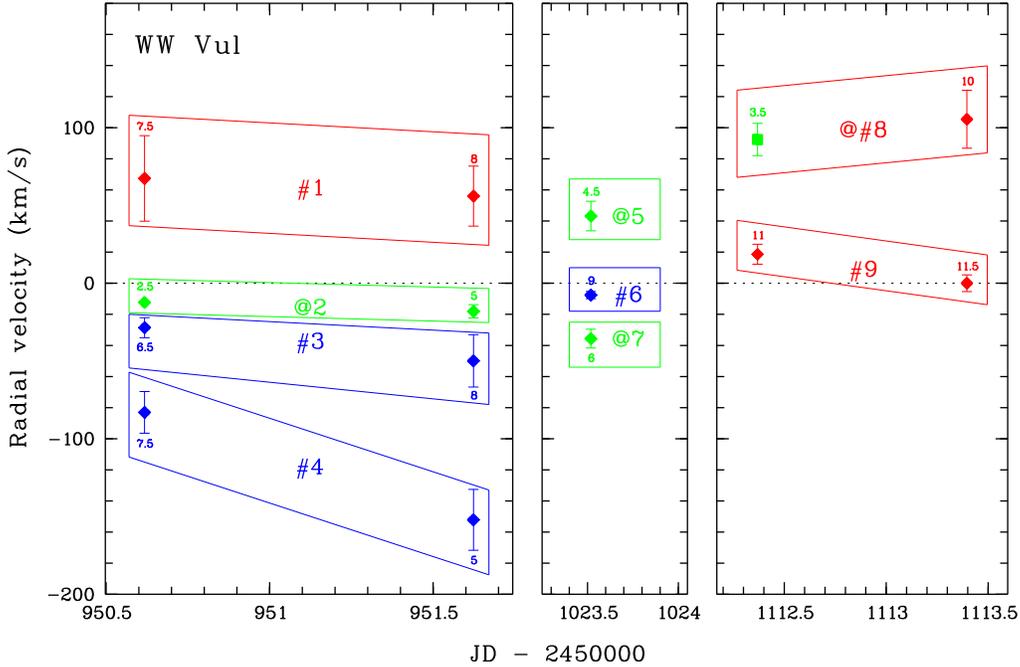}}
\caption{
Events in WW~Vul.
Events marked with a ``\#'' (1, 3, 4, 6, 8 and 9; red for redshifted events and blue for blueshifted events) are detected both in hydrogen and metallic lines.
Events denoted with an ``@'' (2, 5 and 7; green) are only seen in metallic lines.
The square point of \#8 in JD~1112.37 correspond to TACs only observed in \ion{Fe}{ii} and \ion{Ti}{ii}.
Further details as for Fig~\ref{master_bfori}.
(This figure is available in color in electronic form).}
\label{master_wwvul}
\end{figure*}

\begin{figure*}
\centerline{
\includegraphics[clip=true,angle=-90,width=0.9\hsize]{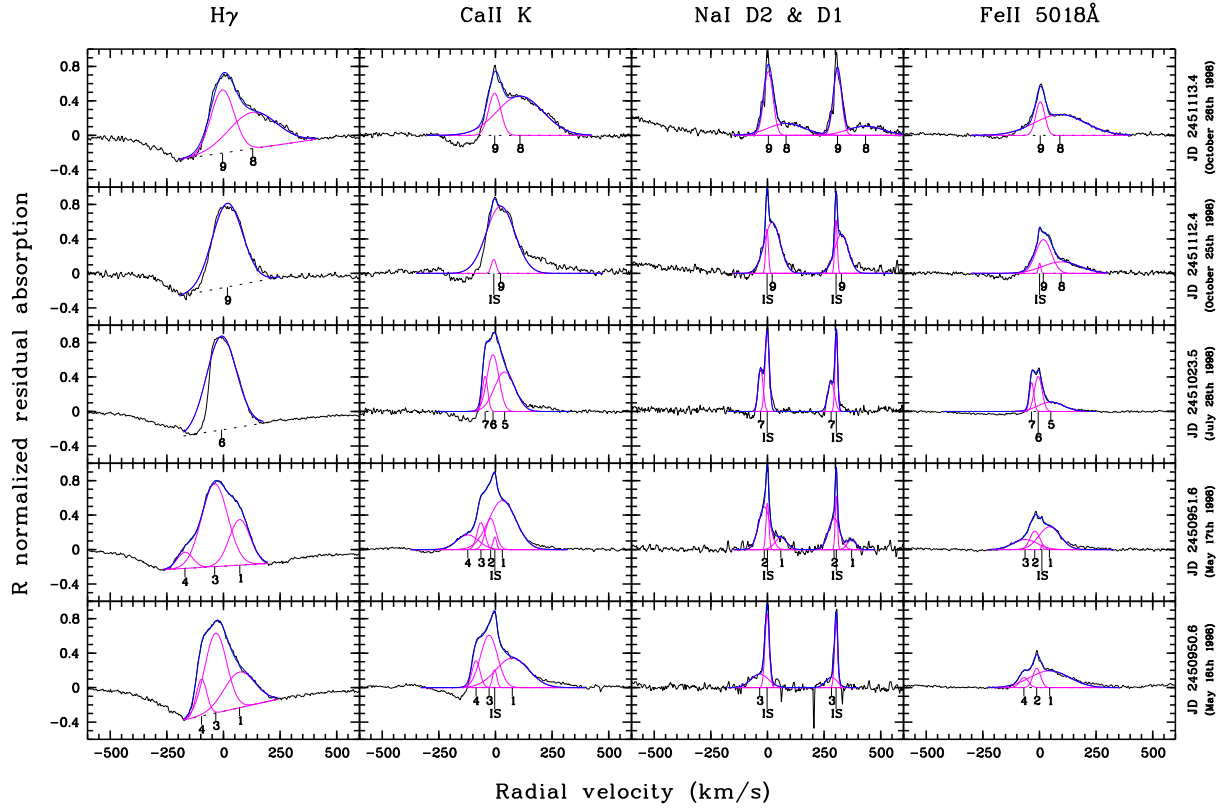}}
\caption{
WW~Vul $R$ profiles.
Details as for Fig~\ref{bfori_spectra}.
(This figure is available in color in electronic form).}
\label{wwvul_spectra}
\end{figure*}

\begin{figure*}
\centerline{
\includegraphics[clip=true,angle=-90,width=0.75\hsize]{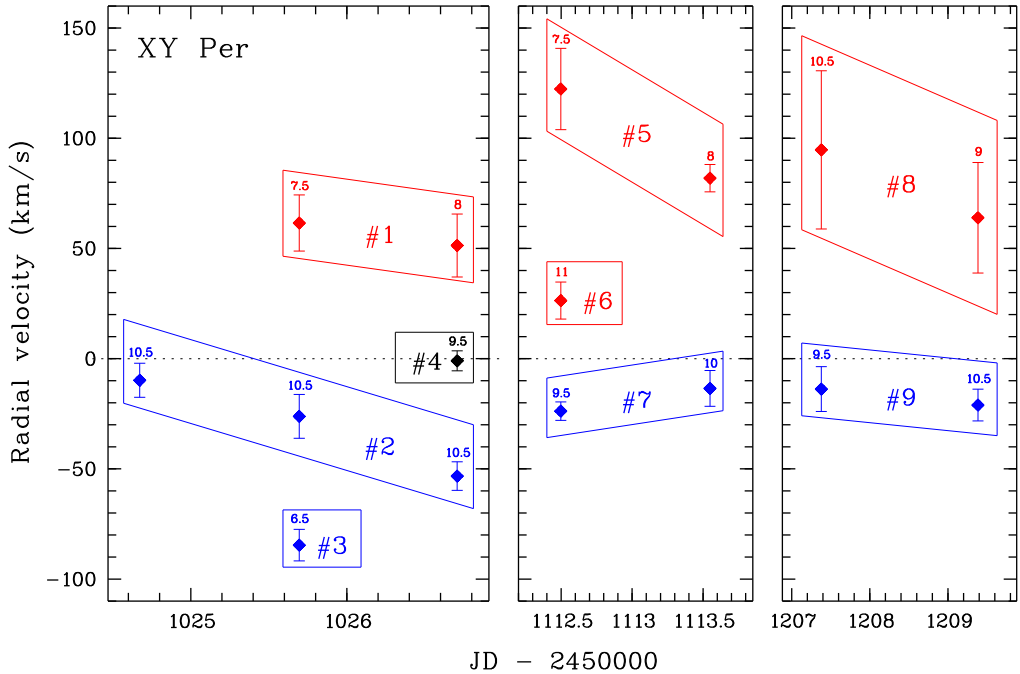}}
\caption{
Events in XY~Per.
Details as for Fig~\ref{master_bfori}.
(This figure is available in color in electronic form).}
\label{master_xyper}
\end{figure*}

\begin{figure*}
\centerline{
\includegraphics[clip=true,angle=-90,width=0.9\hsize]{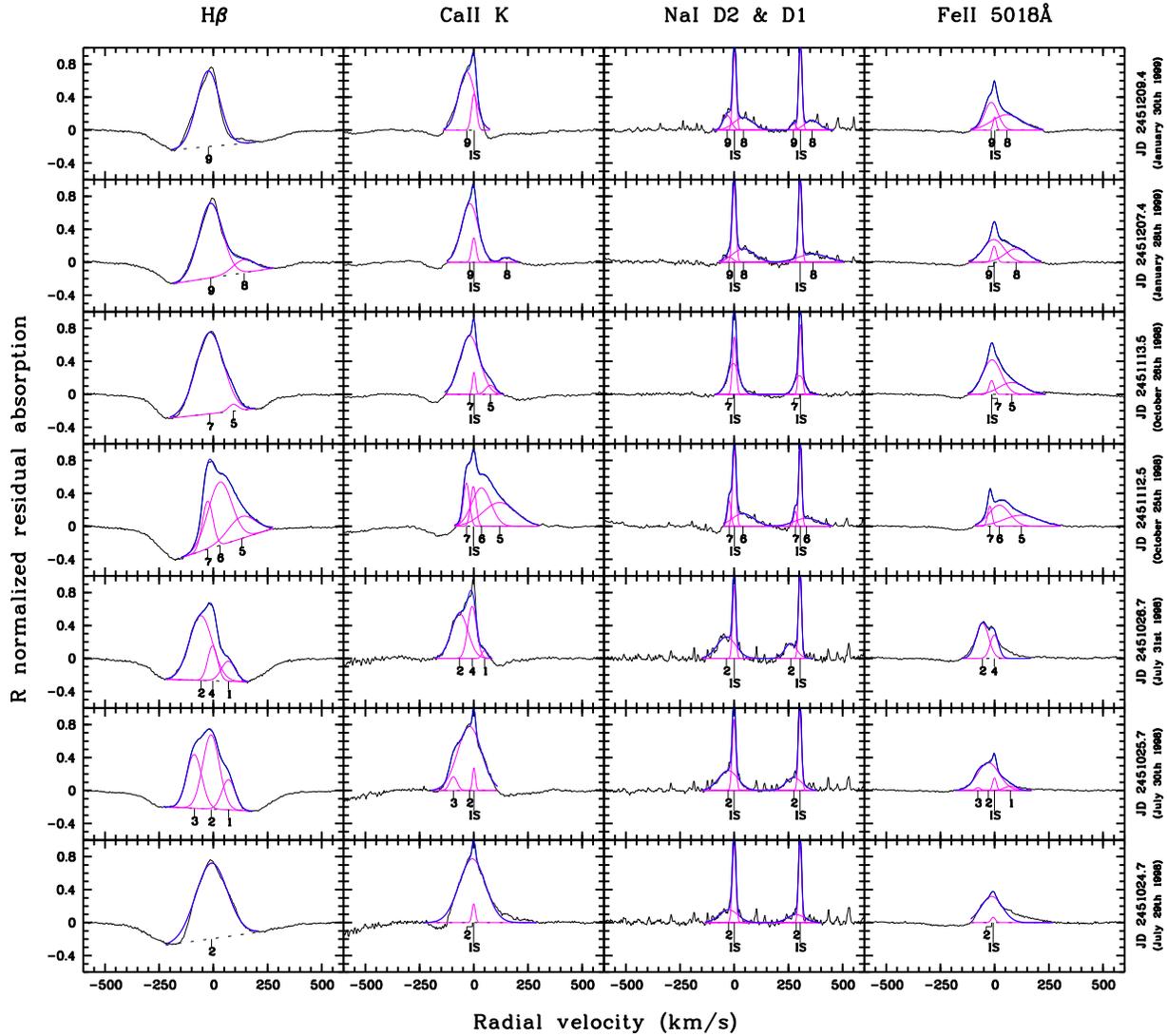}}
\caption{
XY~Per $R$ profiles.
Details as for Fig~\ref{bfori_spectra}.
(This figure is available in color in electronic form).}
\label{xyper_spectra}
\end{figure*}

Tables~\ref{master_table_bfori},~\ref{master_table_svcep},~\ref{master_table_wwvul}~and~\ref{master_table_xyper} show that absorption components with similar radial velocities appear/disappear simultaneously in different lines, as is observed in UX~Ori (\citetalias{mora2002}).
We assume that absorptions with similar velocities come from the same gaseous clump, which can be characterized by an average radial velocity $<$$v$$>$ (a Transient Absorption Component or TAC).
The time evolution of the TACs' velocity is referred to as an event and represents the dynamical evolution of the gaseous clumps.
We point out that there is an uncertainty in identifying TACs detected on different nights with the same gaseous clump; our assumption relies on the UX~Ori results \citepalias[see also below]{mora2002}.
Figs.~\ref{master_bfori},~\ref{master_svcep},~\ref{master_wwvul}~and~\ref{master_xyper} plot $<$$v$$>$  of the identified TAC versus JD.
$<$$v$$>$ is a weighted average in which the lines H$\beta$, H$\gamma$, H$\delta$, \ion{Ca}{ii}~K, \ion{Fe}{ii}~4924~\AA\ and \ion{Fe}{ii}~5018~\AA\ have an arbitrarily assigned weight of 1 because of their higher intensity and non-blended nature.
A weight of 1/2 is also arbitrarily assigned to the remaining lines in order to reflect that they are either blended (H$\epsilon$, \ion{Ca}{ii}~H and \ion{Fe}{ii}~5169\AA), weak (\ion{Sc}{ii}~4247\AA, \ion{Fe}{i}~4046\AA\ and \ion{Ca}{i}~4227\AA) or affected by telluric lines (\ion{Na}{i} D2 \& D1).
The weighted number of lines used in each average, which can be fractional because of the 1/2 weights, is plotted next to each point.
The TACs have been grouped according to the event they represent; thus, the figures show the dynamical evolution of the gaseous clumps.
Note that some events were only detected once (only one TAC).
 Figs.~\ref{bfori_spectra},~\ref{svcep_spectra},~\ref{wwvul_spectra}~and~\ref{xyper_spectra} show the $R$ profiles of some selected lines for the four stars.
The line absorption components and the corresponding  event identifications are indicated.
The lines are H$\beta$, \ion{Ca}{ii}~K (except for BF~Ori), \ion{Na}{i}~D2, \ion{Na}{i}~D1 and \ion{Fe}{ii}~5018~\AA.
H$\gamma$ is shown for WW~Vul (Fig.~\ref{wwvul_spectra}) since H$\beta$ has a large underlying emission.
The results for each star are presented in the following.

{\it BF~Ori:}
5 TACs grouped in 3 different events were detected in the spectra of BF~Ori (Fig.~\ref{master_bfori}).
Event \#1 is an accelerating redshifted event, first detected in the JD~1112.63 spectrum at approximately the stellar radial velocity.
Absorbing gas is seen in Balmer and metallic lines of \ion{Na}{i}, \ion{Fe}{ii}, \ion{Ti}{ii}, \ion{Sc}{ii}, and \ion{Ca}{i}.
The strongest \ion{H}{i} lines (H$\beta$, H$\gamma$ and H$\delta$) appear to be saturated, i.e. $R_{\rm max}$ is very close to unity, and the metallic lines are also very strong.
The Balmer lines are broader than the metallic ones.
The parameters of at least some of these lines might be influenced by IS gas absorption (a careful look at the \ion{Na}{i}~D lines shows the presence of two peaks in the JD~1113.65 spectrum, Fig.~\ref{bfori_spectra}).
The velocity dispersion of the lines tends to be larger when the event increases its velocity, while $R_{\rm max}$ values tend to decrease, though changes are modest.
Event \#2 is a strong redshifted decelerating event; the behaviour of its TACs  is in general similar to those of \#1.
\#3 represents blueshifted gas only detected in the last night of January~99 and is fainter than the redshifted ones.
The $R_{\rm max}$ values of the Balmer lines are low, but they might be saturated, because $R_{\rm max}$ does not appear to decrease as we follow the series in what will be called the ``expected Balmer decreasing trend'' for optically thin gas ($R_{\rm max}($H$\beta) > R_{\rm max}($H$\gamma) > R_{\rm max}($H$\delta) > R_{\rm max}($H$\zeta)$); also the relative intensity of the metallic lines with respect to the \ion{H}{i} ones seems to be lower.

The \ion{Ca}{ii}~K line has non-photospheric profiles with the simultaneous presence of redshifted and blueshifted components, but their radial velocities do not match  the absorptions observed in other lines (except the blueshifted TAC in JD~1210.45).

{\it SV Cep:}
Broad absorptions of \ion{H}{i}, \ion{Ca}{ii}, \ion{Na}{i} and \ion{Fe}{ii} are detected in the spectra of SV~Cep, but no variability is found in \ion{Ti}{ii}, \ion{Sc}{ii}, \ion{Fe}{i} and \ion{Ca}{i} (unlike the other stars in the paper).
The broad absorptions represent 10 TACs grouped in 8 different events: 3 of them correspond to outflowing gas and the remaining 5 to infalling gas (Figs.~\ref{master_svcep}~and~\ref{svcep_spectra}).
Blueshifted gas shows small radial velocities, on average $ \leq 20$~\kms , while redshifted components display velocities as high as 160~\kms.
Only one spectrum was taken in May~98, October~98 and January~99, i.e. the events of these periods are composed of 1 TAC only: these data represent isolated snapshots of the CS gas around SV~Cep and no temporal evolution can be inferred from them.
The 4 TACs detected in July~98 can be grouped in two events: \#3 corresponds to redshifted gas with practically constant radial velocity, and \#4 is gas observed at a velocity close to the stellar radial velocity or slightly blueshifted.
In general, the $R_{\rm max}$ values of the Balmer lines show the expected Balmer decreasing trend and are broader and much stronger than the metallic lines, $R_{\rm max}$(H$\delta$)\,/\,$R_{\rm max}$$(\ion{Fe}{ii})>3$.
There are, however, some exceptions.
\#1 shows relatively strong \ion{Fe}{ii} and it is not clear that the Balmer lines show the expected decreasing trend.
In \#5 the \ion{H}{i} lines are not broader than the \ion{Fe}{ii} ones, and the Balmer lines are probably saturated (but note that this event is very faint, and it could be an overinterpretation of the fit procedure).
The strongest \ion{H}{i} lines are saturated in \#7 and also in the JD~1026.66 spectrum of \#3.
There are anticorrelated changes in the \deltav\ and $R_{\rm max}$ values of \#3, but they show the same trend in \#4.

{\it WW Vul:}
15 TACs grouped in 9 different events are identified in the 5 spectra of WW~Vul (Figs.~\ref{master_wwvul}~and~\ref{wwvul_spectra}). 
4 events are seen in May 98. \#1 is redshifted gas with saturated Balmer lines and strong \ion{Fe}{ii} lines.
Both \ion{H}{i} and \ion{Fe}{ii} lines have similar \deltav\ and from one TAC to another $R_{\rm max}$ and \deltav\  show opposite trends.
\#2 corresponds to low velocity blueshifted gas clearly detected in the metallic lines but no counterpart in the hydrogen lines is apparent (see the \ion{Fe}{ii} 5018~\AA\ line in Fig.~\ref{wwvul_spectra}).
\#3 and \#4 are blueshifted accelerating events.
The only spectrum of July 98 (JD~1023.51) reveals 3 TACs: \#5 is a broad, $\Delta v > 100$~\kms  , redshifted event only detected in metallic lines (the broad wing of the \ion{Fe}{ii} 5018~\AA\ line profile in Fig.~\ref{wwvul_spectra}).
\#6 is a very low velocity blueshifted component \mbox{($v \simeq -5$~\kms)}; this event is significantly broader in the \ion{H}{i} lines (which are  saturared) than in the metallic ones (the IS contribution cannot be separated from this low velocity event).
\#7 is a relatively narrow blueshifted event only detected in metallic lines and is clearly distinguished as a peak in the line profiles.
Again, metallic redshifted absorptions at \mbox{$v \simeq 90 $ \kms} without \ion{H}{i} counterpart are detected on JD~1112.36 (October 98).
Similar metallic absorptions are also detected on JD~1113.39, but on this date saturated hydrogen components with basically the same kinematic parameters (including  the velocity dispersion) are present.
We tentatively identify both TACs with the same event, \#8, although we cannot exclude the possibility that the metallic absorptions detected on each date could be due to different gas. 
Finally, \#9 is a strong, redshifted, decelerating event, identified in \ion{H}{i} and in many metallic lines.
In this case, the hydrogen lines are considerably broader than the metallic ones.
The \ion{H}{i} lines seem saturated on the first night, while on the second one the expected Balmer decreasing trend is observable and on both nights the \ion{Fe}{ii} lines are relatively strong.

{\it XY Per:}
16 TACs grouped in 9 different events are detected in the 7 XY~Per spectra.
\#1 is a faint redshifted event detected on the last two (out of three) July~98 nights.
\#2 corresponds to blueshifted accelerating gas; all three TACs of this event are strong and broad.
\#3 corresponds to a relatively narrow and faint blueshifted  event only detected on JD~1025.69. 
\#4 is also a narrow and faint event, centered at approximately the stellar radial velocity, only detected on JD~1026.71; IS absorption could contribute to the metallic $R_{\rm max}$ values.
\#5 represents  decelerating redshifted gas. 
\#6 is a strong redshifted event only detected on JD~1112.49.
\#7 corresponds to decelerating blueshifted gas; the expected Balmer decreasing trend is clearly present in both TACs and metallic lines are very strong.
Both \deltav\ and $R_{\rm max}$ increase very significantly from JD~1112.49 to JD~1113.55.
\#8 is  decelerating redshifted gas detected on January 98 (note that the time interval between the two TACS grouped in this event is 48 hours); in both TACs the metallic lines are very broad and strong: \ion{Fe}{ii} lines are even broader and stronger than the Balmer ones which are faint and seem saturated.
Finally, \#9 represents accelerating blueshifted gas.    

\section{Discussion}
\label{discussion}

The objects studied in this work and UX~Ori are very much alike.
All are PMS HAe stars. UX~Ori, BF~Ori and WW~Vul are bonafide UXOR-type objects \citep[e.g.][]{grinin2000}, i.e. their light curves show high-amplitude variability ($\Delta m > 2.0$~mag), Algol-like minima, a blueing effect and an increase of the polarization when the object brightness decreases. 
SV~Cep also shows UXOR characteristics \citep{rostopchina2000,oudmaijer2001}.
Further data are required before XY~Per can be confidently classed as a UXOR-type object \citep{oudmaijer2001}, although it does share the same complex and variable spectroscopic behaviour of the other stars.

The time covered by the UX~Ori spectra allowed us to analyze and identify the TACs as due to the dynamical evolution of gaseous moving clumps; this identification and dynamical evolution was particularly convincing in the case of a clump detected in four spectra taken within a time interval of four hours \citepalias{mora2002}.
We rely on those results to ascribe to the same gaseous clump groups of \ion{H}{i} and metallic absorption components described in the previous Sect.
We are aware, however, that this identification is doubtful in some cases, since the number of spectra for the stars in this paper is smaller and the time coverage is poorer than for UX~Ori.
With this caveat in mind, we will discuss these spectra as we have done in \citetalias{mora2002} for UX~Ori.
We are quite confident that the main conclusions of this paper are not affected by the  uncertainties with which some specific event can be identified.

\subsection{Kinematics}
  
Accelerating/decelerating blueshifted and redshifted events were detected in UX~Ori, the events seemed to last for a few days and their acceleration rates were a fraction of a \ms.
Within the present time coverage limitations, the same is observed in the outflowing and infalling gas of BF~Ori, WW~Vul, SV~Cep and XY~Per.
All 5 stars share the trend that infalling gas shows the largest velocities (the exception is event \#4 of WW~Vul in the JD~951.62 spectrum) and that blueshifted absorptions are detected when redshifted absorptions are present (the exception might be the JD~1024.67 spectrum of XY~Per which shows a low velocity outflow but no infalling gas). 
Similar results are also present in the spectra of \citet{grinin2001}. 

Infalling gas appears to have larger dispersion velocities than outflowing gas.
In addition, the \ion{H}{i} lines are  broader than the metallic ones in approximately 40~\% of the identified TACs, as indicated, for example, by the fact that $\Delta v_\ion{Fe}{ii} < 0.66 \Delta v_\ion{H}{i}$ (in UX~Ori this trend was noticed in the \ion{Na}{i} lines).
The fraction of TACs with broader \ion{H}{i} lines varies from star to star.
For example, in 12 out of 16 detected TACs in XY~Per \ion{H}{i} and \ion{Fe}{ii} have similar velocity dispersion, and only 3 out of 10 TACs in SV Sep follow this trend.
On  the other hand, the \ion{Fe}{ii} lines are broader than the \ion{H}{i} lines in only two events, \#5 of SV~Cep and \#8 of XY~Per. 
It is worth noting that the event of XY~Per has metallic $R_{\rm max}$ values at least as large as those of the Balmer lines and that the SV~Cep event is very weak (see above for this event).

There seems to be a correlation between the dispersion velocity and the velocity of the TACs in the sense that TACs appear to be broader when the velocity increases.
If we perform a linear regression ($\Delta v$~=~A~+~B~$\times \: |v|$) to the whole set of data, we find a correlation coefficient of 0.68.
There is also a suggestion of an anticorrelation between $<$$\Delta v_{\ion{Fe}{ii}}$$>$ and the $<R_{\rm max}>$ values of the TACs during their evolution, i.e. events become fainter when they increase their dispersion velocity.
This was also suggested in the case of UX~Ori.
For every event with more than 1 observation (14 in total), $<$$\Delta v_{\ion{Fe}{ii}}$$>$ and $<R_{\rm max}>$ have been normalized to the first observed values.
If 2 `anomalous' events, which are the most uncertain identifications in the whole sample, are removed  (WW~Vul \#8 -- one TAC is observed only in the 
metallic lines while the second TAC is detected in both metallic and \ion{H}{i} lines --, and XY~Per \#7 --  the $\Delta v$ and $<R_{\rm max}>$ variations 
are much more extreme than those of all other detected events) we find a linear correlation coefficient of -0.66.
Both correlations are highly significant, i.e., the probability of randomly obtaining such coefficients from 2 unrelated variables is $<$1\% for the data sets considered. We have to point out, however, that the  $<$$\Delta v_{\ion{Fe}{ii}}$$>$ vs $<R_{\rm max}>$ anticorrelation dissapears if 
the two `anomalous' events are included in the statistics.

\subsection{Line intensity ratios}

Many absorption components in Tables~\ref{master_table_bfori}~to~\ref{master_table_xyper} have $R_{\rm max}$ values close to unity, which suggests that they are saturated.
This is not the case even for the strongest events of UX~Ori \citepalias{mora2002} (in fact, the ratio among the $R_{\rm max}$ line values of the 24 UX~Ori TACs does not vary much, which allowed us to estimate a line residual absorption average). 
Nevertheless, we have followed the procedure of \citetalias{mora2002} to investigate whether a ``fixed'' ratio among the line absorption strengths might be present in those TACs which are most likely unsaturated.
Thus, we have excluded from this exercise lines wich show signs of being saturated, i.e. those with $R_{\rm max} >$~0.8 and TACs with Balmer lines of similar strength.
The line \ion{Fe}{ii}~4924~\AA ~has been taken as a reference as it shows the lowest statistical errors (other lines which have been considered are H$\zeta$, \ion{Ca}{ii}~K and \ion{Fe}{ii}~5018~\AA).
The ratio $R_{\rm max,\,line}/R_{\rm max,\,ref.line}$ (e.g. $R_{\rm max,\,H\delta}/R_{\rm max,\,\ion{Fe}{ii} 4924 \AA}$) has been computed for each line of every TAC, and later the mean $<$$R_{\rm max,\,line} / R_{\rm max,\,ref.line}$$>$ has been estimated using a sigma-clipping algorithm to reject bad points.
Table~\ref{ratios} gives $<$$R_{\rm max,\,line} / R_{\rm max,\,ref.line}$$>$ for each star, together with statistical errors and line rejections (\%).

\begin{table*}
\caption{
Ratios of the average $R_{\rm max}$ parameter of several lines to \ion{Fe}{ii}~4924~\AA\ $<R_{\rm max,\,line} / R_{\rm max,\,\ion{Fe}{ii} 4924 \AA}>$) for each star.
The values correspond to lines in TACs which are most likely not saturated (see text).
Values with no error mean that only one TAC is available.
The sigma-clipping threshold adopted is 2.0~$\sigma$, except for the values followed by the symbol $\dagger$, in which 1.5~$\sigma$ has been used.
The percentage of rejected lines is given in brackets.}
\label{ratios}
\centerline{
\begin{tabular}{lllll}
\hline
\hline
Line                 & BF Ori                 & SV Cep                 &
                       WW Vul                 & XY Per                 \\
\hline
H$\beta$             &  --                    & 5.23 $\pm$ 1.82        &
                       6.59                   & 3.40 $\pm$ 0.83 (14\%) \\
H$\gamma$            &  --                    & 5.38 $\pm$ 2.29        &
                       4.53                   & 2.68 $\pm$ 0.82 (11\%) \\
H$\delta$            &  --                    & 4.96 $\pm$ 1.85        &
                       4.83 $\pm$ 0.97        & 2.86 $\pm$ 1.15        \\
H$\zeta$             & 1.50 $\pm$ 0.30        & 3.71 $\pm$ 1.27        &
                       2.17 $\pm$ 0.69        & 2.09 $\pm$ 0.66        \\
\ion{Ca}{ii} K       & 5.79                   & 4.05 $\pm$ 0.97 (10\%) &
                       2.37 $\pm$ 0.65        & 2.20 $\pm$ 1.16 (8\%)  \\
\ion{Na}{i} D2       & 1.06 $\pm$ 0.40        & 1.27 $\pm$ 0.40        &
                      1.45 $\pm$ 0.64 (17\%)$^\dagger$&0.76 $\pm$ 0.12 (20\%)\\
\ion{Na}{i} D1       & 0.89 $\pm$ 0.39        & 0.85 $\pm$ 0.33        &
                       1.50 $\pm$ 0.92        & 0.48 $\pm$ 0.09 (11\%) \\
\ion{Fe}{ii} 5018\AA & 1.08 $\pm$ 0.15        & 1.41 $\pm$ 0.14        &
                       1.28 $\pm$ 0.11 (15\%) & 1.16 $\pm$ 0.17 (14\%) \\
\ion{Fe}{ii} 5169\AA & 1.18 $\pm$ 0.15        & 1.52 $\pm$ 0.30        &
                       1.37 $\pm$ 0.13 (15\%) & 1.30 $\pm$ 0.25 (7\%)  \\
\ion{Ti}{ii} 4444\AA & 0.57 $\pm$ 0.29        & --                     &
                       0.25 $\pm$ 0.08        & 0.30 $\pm$ 0.14 (7\%)  \\
\ion{Ti}{ii} 4572\AA & 0.45 $\pm$ 0.07        & --                     &
                       0.25 $\pm$ 0.08        & 0.32 $\pm$ 0.10 (7\%)  \\
\ion{Sc}{ii} 4247\AA & 0.47 $\pm$ 0.09        & --                     &
                       0.31 $\pm$ 0.06        & 0.33 $\pm$ 0.23        \\
\ion{Ca}{i}  4227\AA & 0.32 $\pm$ 0.06        & --                     &
                       0.18 $\pm$ 0.11        & 0.26 $\pm$ 0.08        \\
\ion{Fe}{i}  4046\AA &  --                    & --                     &
                       0.16 $\pm$ 0.06        & 0.18 $\pm$ 0.05        \\
\hline
\end{tabular}}
\end{table*}

Values quoted in Table~\ref{ratios} can be used to assess whether the gas causing the variable absorptions is optically thin or not, simply by comparing the $R_{\rm max}$ ratios of lines belonging to the same element multiplet with their gf ratios (we recall that the TACs suspected of being saturated have already been excluded).
We have analysed the Balmer, the \ion{Na}{i}~D doublet and the \ion{Fe}{ii} 42 lines, for which H$\delta$, \ion{Na}{i}~D2 and \ion{Fe}{ii}~5018~\AA\ have been taken as reference, respectively.
$R_{\rm max}$ ratios have been computed following the above procedure and gf values have been taken from the VALD database \citep{kupka1999} for \ion{H}{i} and \ion{Na}{i}.
\ion{Fe}{ii} lines do not have reliable experimental gf values, partly because \ion{Fe}{ii}~5169~\AA\ is blended with \ion{Mg}{i}~5167~\AA, \ion{Fe}{i}~5167~\AA\ and \ion{Mg}{i}~5173~\AA. Following the suggestion made by T.A.~Ryabchikova and F.G.~Kupka (private communication), we have used the semiempirical values computed according to \citet{raassen1998} and available at {\it http://www.science.uva.nl/research/atom/levels/levtext.html}.
The comparison between the estimated ratios and the theoretical gf shows that the \ion{Fe}{ii} 42 triplet is most likely saturated in BF~Ori but not in the rest of the stars, and that the \ion{Na}{i}~D doublet is also probably saturated in BF~Ori and WW~Vul (Table~\ref{multiplets}).
Concerning the Balmer lines (note we are referring to those TACs apparently unsaturated) their ratios are very different from the  gf ratios.
This is, in principle, similar to the case of UX~Ori where the lines do not seem to be saturated in any of its events.
\citetalias{mora2002} suggested that the UX~Ori results could be explained by underlying line emission caused by a spherical occulting cloud with a temperature T$_{\rm ex} \sim$~7000~K and a radius of the order of the UX~Ori corotation radius, $R_{\rm cloud}/R_{*} \sim$~1.6 \citep[see][for details of the assumptions]{rodgers2002}.
However, only XY~Per presents Balmer line ratios which could be adjusted using this scenario, namely gas at a T$_{\rm ex} \sim$~6600 K at approximately the corotation radius $R_{\rm cloud}/R_{*} \sim$~1.6.
This is not applicable for the rest of the stars, where it is most likely that the Balmer lines are always saturated.

\begin{table*}
\caption{Estimated $<R_{\rm max,\,line} / R_{\rm max,\,ref.line}>$ ratios among the lines of the \ion{Na}{i}~D doublet and the \ion{Fe}{ii} 42 triplet for each star.
The theoretical ratios (gf$_{\rm line}$~/~gf$_{\rm ref.line}$) and the line taken as reference are given in the last Col.
The \ion{Fe}{ii}~5169~\AA\ line is included for comparison purposes though its ratios are likely affected by the blend with \ion{Mg}{i}~5167~\AA, \ion{Fe}{i}~5167~\AA\ and \ion{Mg}{i}~5173~\AA.}
\label{multiplets}
\centerline{
\begin{tabular}{lllllll}
\hline
\hline
Line                 & BF Ori          & SV Cep          & WW Vul          &
                       XY Per          & Theor. & Reference            \\
\hline
\ion{Na}{i} D1       & 0.83 $\pm$ 0.11 & 0.73 $\pm$ 0.15 & 0.81 $\pm$ 0.12 &
                       0.63 $\pm$ 0.07 & 0.50   & \ion{Na}{i} D2       \\
\ion{Fe}{ii} 4924\AA & 0.94 $\pm$ 0.16 & 0.72 $\pm$ 0.08 & 0.79 $\pm$ 0.07 &
                       0.86 $\pm$ 0.17 & 0.69   & \ion{Fe}{ii} 5018\AA \\
\ion{Fe}{ii} 5169\AA & 1.09 $\pm$ 0.03 & 1.07 $\pm$ 0.15 & 1.06 $\pm$ 0.13 &
                       1.07 $\pm$ 0.19 & 1.25   & \ion{Fe}{ii} 5018\AA \\
\hline
\end{tabular}}
\end{table*}

\subsection{Origin of the variable circumstellar gas clumps detected in hydrogen and metallic lines}

Excluding the fact that redshifted events seem to be detected at larger velocities than blueshifted events (a fact deserving further observations and also a theoretical explanation), the gas is observed in the same absorption lines with similar $R$ and 
velocity dipsersions, i.e., there are no fundamental differences in the behaviour of the infalling and outflowing gas in all the observed stars, including UX~Ori. This result suggests that their physical conditions are 
rather similar and that they probably originate at similar distances from the 
star.
Roughly solar abundances were found by \citet{natta2000} for an event of redshifted gas in UX~Ori, which was further supported by \citetalias{mora2002}.
This led to the conclusion that the clumps of CS gas in UX~Ori are non metal-rich.
Since high velocity gas is observed simultaneously in the Balmer and metallic 
lines in BF~Ori, SV~Sep, XY~Per and 
in most of the detected events in WW~Vul, a similar conclusion very likely holds also for the CS gas in these stars (but see next Sect. for the metallic events in WW~Vul).
Unlike the detected events in UX~Ori, the gas in those stars is often optically thick, as suggested by the saturation of the absorption features.
This might indicate that high density gas is more frequently observed in  BF~Ori, SV~Sep, WW~Vul and XY~Per than in UX~Ori.     
 
\citetalias{mora2002} compares the dynamics of the gaseous clumps in UX~Ori with the predictions of magnetospheric accretion models \citep[see][for a very good basic description of this theory]{hartmann1998} and different wind models \citep[e.g.][]{goodson1997,shu2000,konigl2000}.
The present data  do not add any new substantial aspect to that discussion, only that, with very few exceptions, outflowing gas displays smaller velocities than infalling gas.
Thus, to avoid repetition we refer to that paper, stressing the need for further theoretical efforts to explain the complex circumstellar environment of HAe stars, at least to the level of understanding achieved for the less massive T~Tauri stars.

\subsection{The intriguing case of WW Vul}

\begin{figure}
\centerline{
\includegraphics[clip=true,angle=-90,width=1.0\hsize]{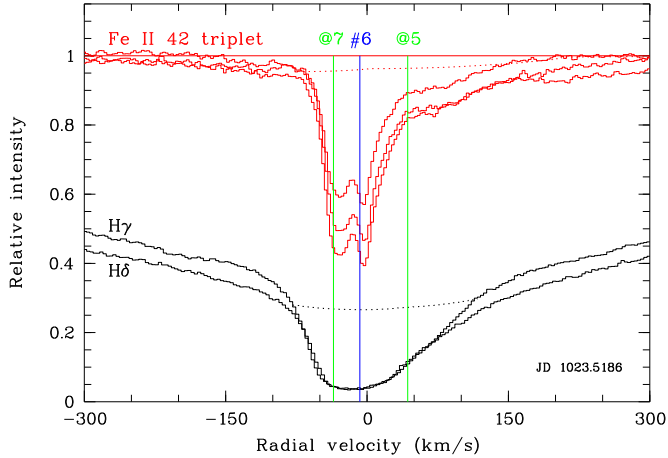}}
\caption{
Original spectrum (i.e. prior to the synthetic stellar spectrum subtraction) 
as observed in the JD~1023.52 spectrum of WW~Vul.
Wavelengths have been converted into radial velocities. 
Plotted lines are H$\gamma$, H$\delta$ and the three lines of the 
\ion{Fe}{ii} 42 triplet.
The photospheric synthetic spectrum used to form $R$ is shown in dotted lines for H$\gamma$ and \ion{Fe}{II}~4924\AA\ (being very similar for the other lines).
The synthetic spectra reveal the presence of emission in the wings of H$\gamma$ and a little in the blue wing of \ion{Fe}{II}~4924\AA.
The average radial velocity $<$$v$$>$ of the events identified after the photospheric spectrum removal and $R$ multigaussian fit is indicated using the same colour and label conventions of Fig.~\ref{master_wwvul}.
There is 1 blueshifted TAC (\#6, blue colour) observed both in metallic and hydrogen lines, note that $\Delta v_{\ion{H}{i}} \gg \Delta v_{\ion{Fe}{ii}}$.
There are also 2 metallic TACs (@5, @7, green) observed only in metallic lines.
Event @7 can be clearly identified by visual inspection as a relatively narrow 
peak in the \ion{Fe}{ii} lines.
In events \#6 and @7 $<$$v$$>$ does not exactly correspond with the local minima in the \ion{Fe}{ii} spectrum because the whole $R$ profile is used in the multigaussian fits.
(This figure is available in color in electronic form).}
\label{wwvul_metallic}
\end{figure}

WW~Vul is a very interesting case, somewhat different from the other stars.
In all the spectra we have obtained, in addition to events detected in both metallic and hydrogen lines, as it is always the case in the other objects, we see also metallic absorption features (both blueshifted and redshifted) that do not seem to have a counterpart in the hydrogen lines.
They appear as broad high velocity wings in the $R$ profiles, e.g. events \#5 and \#8 in the JD 1023.52 and 1112.37 spectra, as well as relatively narrow distinct peaks, e.g. \#7 in JD~1023.52 (Fig.~\ref{wwvul_spectra}).
The observed outflowing events are narrower and have larger $R_{\rm max}$ values than the infalling ones. 
To ensure that the metallic events are not artificially introduced by the multigaussian fit we have proceeded in two different ways:
1. We have carefully inspected the original spectra, i.e., before subtraction of the photospheric components.
Fig.~\ref{wwvul_metallic} shows several lines as observed in the original (without photospheric subtraction) JD~1023.52 WW~Vul spectrum.
In particular, the relatively narrow, blueshifted, metallic event \#7 is clearly seen on a simple visual inspection as an absorption peak with no distinctly separate hydrogen counterpart.
We do note that there is hydrogen absorption covering the velocity range of event \#7 and that the hydrogen lines are quite broad and likely to be saturated at this time.
As a result, Gaussian fitting may not properly represent the intrinsic shape of the hydrogen components; however, without advance knowlegde of their intrinsic shape, Gaussian fitting is used here as a default procedure.
2. In general, our approach has been to use a number as small as possible of gaussians to reproduce the $R$ profiles in a self-consistent way, avoiding to overfit the data.

Nevertheless, since we were intrigued by the WW Vul behaviour, we have performed a number of additional numerical tests, to check as well as possible that the apparent lack of a hydrogen counterpart to some event seen in the metallic lines was not an artifact.
First of all, we have added to the \ion{H}{i} identified components more gaussians which take into account the parameters of the absorptions only identifed in the metallic lines.
This led to unphysical results, i.e., adding more gaussians to the \ion{H}{i} lines does not produce kinematic components similar to those observed in the metallic lines, but create spurious \ion{H}{i} components.
For example, in JD~1023.52 we tried to fit 3 components to the \ion{H}{i} $R_{\rm max}$ profiles, i.e. the same number of gaussians obtained from the \ion{Fe}{ii} fit.
The data obtained from the Balmer lines deconvolution are not self-consistent:
i) strange, unrealistic Balmer intensity sequences are obtained, e.g. in event \#7 $R_{\rm max, H\zeta} = R_{\rm max, H\beta} > R_{\rm max, H\gamma} > R_{\rm max, H\delta}$.
ii) the intensity ratios between different events differ widely from line to line (e.g. H$\beta$ is about two times more intense in \#5 than in \#6, H$\gamma$ is only 0.8 times as intense in \#5 as in \#6.
We have then also tried to fix the radial velocity of the gaussian \ion{H}{i} components to the values obtained from the \ion{Fe}{ii} lines in order to improve the fit, but the results were even worst: the method diverged for H$\beta$ (i.e. no fit could be found) and the sequence of intensities for the Balmer series in event \#6 was unphysical ($R_{\rm max, H\zeta} = R_{\rm max, H\delta} = 2.5  R_{\rm max, H\gamma}$, no H$\beta$ component).

One can also argue that the absence of \ion{H}{i} counterparts to the metallic events is a consequence of blending, which is unresolved by the multigaussian deconvolution procedure.
In order to test such possibility, we have generated a composite synthetic R profile consisting of three gaussians with the same $R_{\rm max}$ and $v$ as the \ion{Fe}{II}~5018 \AA ~JD 1023.52 spectrum and $\Delta v$ = 150 km/s, which is the average velocity dispersion of the only \ion{H}{i} component detected in that date.
We have also added random noise to achieve a S/N ratio worst than that of the H$\beta$ and H$\gamma$ lines of that spectrum; finally, we have applied our multigaussian deconvolution procedure to the synthetic profile.
The result is that the three individual gaussians have been successfully retrieved.
Note that the gaussian parameters we have introduced are test values.
This result suggests that we would have been able of identifying \ion{H}{i} counterparts of the metallic kinematics components if they would exist. Therefore, we consider that the existence of metallic absorptions without obvious  \ion{H}{i} counterparts is a rather firm result.

There are no obvious differences, in terms of kinematics and absorption strength, between the  WW~Vul metallic events and the rest of the events detected in this star and in BF~Ori, SV~Cep, XY~Per and UX Ori. Some remarkable differences appear, however, when the ion column densities causing the events are compared.
Lower limits on the ion column densities causing the absorptions can be estimated according to the following formula \citep{spitzer1978}, which becomes exact when the gas is optically thin:
\begin{equation}
\label{eqn_column_densities}
N_a \simeq 
\frac{4 \pi \epsilon_0}{e^2} \frac{m_e c^2 W_\lambda}{\pi f \lambda^2}
\end{equation}

Where $N_a$ is the column density of ions in the ground state of the line, $e$ is the elementary charge, $m_e$ is the electron mass, $\epsilon_0$ is the permeability of free space, $c$ is the speed of light, $f$ is the oscillator strength of the line, $\lambda$ is the wavelength of the line and $W_\lambda$ is the equivalent width of the absorption component.
The column densities obtained are similar, within an order of magnitude, for all the metallic events.
An upper limit on the column density of the \ion{H}{i} atoms in the Balmer energy level (n=2) can be estimated if we assume that TACs with an intensity lower than 3 times the noise level of the spectra cannot be detected.
Table~\ref{column_densities} gives the estimated values for the \ion{Fe}{ii} lines and some Balmer lines in the case of event WW Vul \#7.
Event \#8 in XY~Per has extremely weak saturated \ion{H}{i} features and strong metallic components (\ion{H}{i} line strengths  are lower than or just comparable to the \ion{Fe}{ii} lines); in  this sense, this event is the more similar to the metallic ones among all detected events with \ion{H}{i} and metallic components.
Table~\ref{column_densities} gives lower limits to the column densities, estimated from the Balmer and \ion{Fe}{ii} lines of event \#8 in XY~Per. 
The values of Table~\ref{column_densities} nicely shows the difference between the estimated column densities of ions excited to the energy ground level of the \ion{Fe}{ii} 42 triplet and Balmer lines in both type of events, and it could point out to a fundamental difference on the nature and origin of the gas from which the absorptions rise.
However, a definitive statement on this issue requires a deep analysis of the metallic events detected in WW Vul in order to estimate chemical abundances, or abundance ratios among elements.
Such analysis, which is beyond the scope of this paper, needs a NLTE treatment of the spectra; such treatment  would produce model dependent results, since a previous knowledge of physical quantities, as for example volume densities of the gaseous clumps and electron temperature, are required.
Such quantities cannot be estimated from our data in a confident way.

\begin{table}
\caption{Column density estimates (according to Eq.~\ref{eqn_column_densities}) of ions excited to the energy ground level of Balmer and \ion{Fe}{ii} 42 triplet lines for events WW~Vul \#7 (metallic) and XY~Per \#8.}
\label{column_densities}
\centerline{
\begin{tabular}{lll}
\hline
\hline
                     & WW Vul \#7             & XY Per \#8             \\
\hline
Line                 & $N_a\,({\rm cm}^{-2})$ & $N_a\,({\rm cm}^{-2})$ \\
\hline
\ion{Fe}{ii} 4924\AA & $\ge1.1 \cdot 10^{14}$ & $\ge4.0 \cdot 10^{14}$ \\
\ion{Fe}{ii} 5018\AA & $\ge8.1 \cdot 10^{13}$ & $\ge2.1 \cdot 10^{14}$ \\
\ion{Fe}{ii} 5169\AA & $\ge7.8 \cdot 10^{13}$ & $\ge4.4 \cdot 10^{14}$ \\
H$\beta$             &   $<2.9 \cdot 10^{11}$ & $\ge1.3 \cdot 10^{13}$ \\
H$\gamma$            &   $<1.2 \cdot 10^{12}$ & $\ge4.2 \cdot 10^{13}$ \\
H$\delta$            &   $<3.3 \cdot 10^{12}$ & $\ge9.6 \cdot 10^{13}$ \\
H$\zeta$             &                        & $\ge2.9 \cdot 10^{14}$ \\
\hline
\end{tabular}
}
\end{table}

\section{Concluding remarks}
\label{conclussions}

We have analyzed optical high resolution spectra of the HAe stars BF Ori, SV Cep, WW Vul and XY Per.
These spectra monitor the stars on time scales of months and days and, as in the case of the previously studied UX Ori \citepalias{mora2002}, they provide observational constraints, which should be considered for any realistic scenario of the gaseous circumstellar disks around intermediate-mass PMS stars.
Our results and conclusions can be summarized as follows:

\begin{enumerate}

\item
The gaseous circumstellar environment of these stars is very complex and active.
The spectra always show circumstellar line absorptions with remarkable variations in their strength and dynamical properties.

\item 
Variable absorption features are, in most  cases, detected simultaneously in hydrogen and in many metallic lines with similar velocities. 
In each case, there are several  kinematic components in each line, both blue-shifted and red-shifted with respect to the systemic velocity, denoting the simultaneous presence of infalling and outflowing gas.
We attribute the variable features detected in both Balmer and metallic lines to gaseous clumps of solar-like composition, evolving dynamically in the circumstellar disks of these objects.
In this respect, the disks around the stars studied in this paper are similar to the UX Ori disk.
Following the conclusions of \citetalias{mora2002} we suggest that these clumps and their dynamical evolution should be investigated in the context of detailed magnetospheric accretion models, similar to those of T Tauri stars.    

\item
The star WW Vul is peculiar and behaves differently from the other stars studied in this paper and also from UX~Ori.
It is the only star that shows, in addition to events seen both in metallic and hydrogen lines, similar to those observed in the other stars, also transient absorption components in metallic lines that do not apparently have any obvious counterpart in the hydrogen lines.
This result, taken at its face-value, would indicate the presence of a metal-rich gas component in the environment of WW~Vul, possibly related to the evaporation of solid bodies. However, any such conclussion is premature.
We think that a series of optical spectra with  better time resolution (hours) and longer monitoring (up to around seven days), spectra in the far UV range - to analyze Lyman and metallic resonance lines - and detailed NLTE models of different CS gas environments are essential for further progress and to provide clues on the origin of these apparently metal-rich  events, in terms of their appearance/disapperance statistics, dynamics, metallicity  and nature. 

\end{enumerate}

\begin{acknowledgements}
The authors wish to thank V.~P.~Grinin for valuable discussion about the analysis procedures and F.~G.~Kupka and T.~A.~Ryabchikova for information about the best gf values for the \ion{Fe}{ii} 42 multiplet.
A. Mora acknowledges the hospitality and support of the Osservatorio Astrofisico di Arcetri for two long stays during which a substantial part of this work was carried out.
A.~Alberdi, C.~Eiroa, B.~Mer\'{\i}n, B.~Montesinos, A.~Mora, J.~Palacios and E.~Solano have been partly supported by Spanish grants ESP98-1339 and AYA2001-1124.
\end{acknowledgements}


\bibliographystyle{aa}
\bibliography{bibtex}

\begin{thebibliography}{30}
\expandafter\ifx\csname natexlab\endcsname\relax\def\natexlab#1{#1}\fi

\bibitem[{Beckwith {et~al.}(2000)Beckwith, Henning, \& Nakagawa}]{beckwith2000}
Beckwith, S. V.~W., Henning, T., \& Nakagawa, Y. 2000, in Protostars and
  Planets IV, ed. V.~Mannings, A.~P. Boss, \& S.~S. Russell (Tucson: University
  of Arizona Press)

\bibitem[{{Beust} {et~al.}(2001){Beust}, {Karmann}, \& {Lagrange}}]{beust2001}
{Beust}, H., {Karmann}, C., \& {Lagrange}, A.-M. 2001, \aap, 366, 945

\bibitem[{{Chen} \& {Jura}(2003)}]{chen2003}
{Chen}, C.~H. \& {Jura}, M. 2003, \apj, 582, 443

\bibitem[{{Eiroa} {et~al.}(2001){Eiroa}, {Garz{\' o}n}, {Alberdi}, {de Winter},
  {Ferlet}, {Grady}, {Cameron}, {Davies}, {Deeg}, {Harris}, {Horne}, {Mer{\'
  i}n}, {Miranda}, {Montesinos}, {Mora}, {Oudmaijer}, {Palacios}, {Penny},
  {Quirrenbach}, {Rauer}, {Schneider}, {Solano}, {Tsapras}, \&
  {Wesselius}}]{eiroa2001}
{Eiroa}, C., {Garz{\' o}n}, F., {Alberdi}, A., {et~al.} 2001, \aap, 365, 110

\bibitem[{{Eiroa} {et~al.}(2002){Eiroa}, {Oudmaijer}, {Davies}, {de Winter},
  {Garz{\' o}n}, {Palacios}, {Alberdi}, {Ferlet}, {Grady}, {Cameron}, {Deeg},
  {Harris}, {Horne}, {Mer{\'{\i}}n}, {Miranda}, {Montesinos}, {Mora}, {Penny},
  {Quirrenbach}, {Rauer}, {Schneider}, {Solano}, {Tsapras}, \&
  {Wesselius}}]{eiroa2002}
{Eiroa}, C., {Oudmaijer}, R.~D., {Davies}, J.~K., {et~al.} 2002, \aap, 384,
  1038

\bibitem[{{Finkenzeller} \& {Jankovics}(1984)}]{finkenzeller1984}
{Finkenzeller}, U. \& {Jankovics}, I. 1984, \aaps, 57, 285

\bibitem[{{Goodson} {et~al.}(1997){Goodson}, {Winglee}, \&
  {Boehm}}]{goodson1997}
{Goodson}, A.~P., {Winglee}, R.~M., \& {Boehm}, K. 1997, \apj, 489, 199

\bibitem[{{Grady} {et~al.}(1996){Grady}, {Perez}, {Talavera}, {Bjorkman}, {de
  Winter}, {The}, {Molster}, {van den Ancker}, {Sitko}, {Morrison}, {Beaver},
  {McCollum}, \& {Castelaz}}]{grady1996}
{Grady}, C.~A., {Perez}, M.~R., {Talavera}, A., {et~al.} 1996, \aaps, 120, 157

\bibitem[{Grady {et~al.}(2000)Grady, Sitko, Russell, {et~al.}}]{grady2000}
Grady, C.~A., Sitko, M.~L., Russell, R.~W., {et~al.} 2000, in Protostars and
  Planets IV, ed. V.~Mannings, A.~P. Boss, \& S.~S. Russell (Tucson: University
  of Arizona Press)

\bibitem[{{Grinin}(2000)}]{grinin2000}
{Grinin}, V.~P. 2000, in ASP Conf. Ser. 219: Disks, Planetesimals, and Planets,
  216

\bibitem[{{Grinin} {et~al.}(2001){Grinin}, {Kozlova}, {Natta}, {Ilyin},
  {Tuominen}, {Rostopchina}, \& {Shakhovskoy}}]{grinin2001}
{Grinin}, V.~P., {Kozlova}, O.~V., {Natta}, A., {et~al.} 2001, \aap, 379, 482

\bibitem[{{Grinin} {et~al.}(1994){Grinin}, {The}, {de Winter}, {Giampapa},
  {Rostopchina}, {Tambovtseva}, \& {van den Ancker}}]{grinin1994}
{Grinin}, V.~P., {The}, P.~S., {de Winter}, D., {et~al.} 1994, \aap, 292, 165

\bibitem[{{Hartmann}(1998)}]{hartmann1998}
{Hartmann}, L. 1998, {Accretion processes in star formation} (Cambridge
  astrophysics series, 32. ISBN 0521435072.)

\bibitem[{{Herbst} \& {Shevchenko}(1999)}]{herbst1999}
{Herbst}, W. \& {Shevchenko}, V.~S. 1999, \aj, 118, 1043

\bibitem[{{K{\" o}nigl} \& {Pudritz}(2000)}]{konigl2000}
{K{\" o}nigl}, A. \& {Pudritz}, R.~E. 2000, Protostars and Planets IV, 759

\bibitem[{{Kupka} {et~al.}(1999){Kupka}, {Piskunov}, {Ryabchikova}, {Stempels},
  \& {Weiss}}]{kupka1999}
{Kupka}, F., {Piskunov}, N., {Ryabchikova}, T.~A., {Stempels}, H.~C., \&
  {Weiss}, W.~W. 1999, \aaps, 138, 119

\bibitem[{{Kurucz}(1993)}]{kuruczCD13}
{Kurucz}, R.~L. 1993, ATLAS9 Stellar Atmosphere Programs and 2 km/s
  grid.~Kurucz CD-ROM No.~13.~ Cambridge, Mass.: Smithsonian Astrophysical
  Observatory, 1993., 13

\bibitem[{Lagrange {et~al.}(2000)Lagrange, Backman, \&
  Artymowicz}]{lagrange2000}
Lagrange, A.-M., Backman, D.~E., \& Artymowicz, P. 2000, in Protostars and
  Planets IV, ed. V.~Mannings, A.~P. Boss, \& S.~S. Russell (Tucson: University
  of Arizona Press)

\bibitem[{{Mora} {et~al.}(2001){Mora}, {Mer{\'{\i}}n}, {Solano}, {Montesinos},
  {de Winter}, {Eiroa}, {Ferlet}, {Grady}, {Davies}, {Miranda}, {Oudmaijer},
  {Palacios}, {Quirrenbach}, {Harris}, {Rauer}, {Cameron}, {Deeg}, {Garz{\'
  o}n}, {Penny}, {Schneider}, {Tsapras}, \& {Wesselius}}]{mora2001}
{Mora}, A., {Mer{\'{\i}}n}, B., {Solano}, E., {et~al.} 2001, \aap, 378, 116

\bibitem[{{Mora} {et~al.}(2002){Mora}, {Natta}, {Eiroa}, {Grady}, {de Winter},
  {Davies}, {Ferlet}, {Harris}, {Montesinos}, {Oudmaijer}, {Palacios},
  {Quirrenbach}, {Rauer}, {Alberdi}, {Cameron}, {Deeg}, {Garz{\' o}n}, {Horne},
  {Mer{\' i}n}, {Penny}, {Schneider}, {Solano}, {Tsapras}, \&
  {Wesselius}}]{mora2002}
{Mora}, A., {Natta}, A., {Eiroa}, C., {et~al.} 2002, \aap, 393, 259

\bibitem[{{Natta} {et~al.}(2000){Natta}, {Grinin}, \&
  {Tambovtseva}}]{natta2000}
{Natta}, A., {Grinin}, V.~P., \& {Tambovtseva}, L.~V. 2000, \apj, 542, 421

\bibitem[{{Natta} {et~al.}(2001){Natta}, {Prusti}, {Neri}, {Wooden}, {Grinin},
  \& {Mannings}}]{natta2001}
{Natta}, A., {Prusti}, T., {Neri}, R., {et~al.} 2001, \aap, 371, 186

\bibitem[{{Oudmaijer} {et~al.}(2001){Oudmaijer}, {Palacios}, {Eiroa}, {Davies},
  {de Winter}, {Ferlet}, {Garz{\' o}n}, {Grady}, {Cameron}, {Deeg}, {Harris},
  {Horne}, {Mer{\' i}n}, {Miranda}, {Montesinos}, {Mora}, {Penny},
  {Quirrenbach}, {Rauer}, {Schneider}, {Solano}, {Tsapras}, \&
  {Wesselius}}]{oudmaijer2001}
{Oudmaijer}, R.~D., {Palacios}, J., {Eiroa}, C., {et~al.} 2001, \aap, 379, 564

\bibitem[{{Raassen} \& {Uylings}(1998)}]{raassen1998}
{Raassen}, A.~J.~J. \& {Uylings}, P.~H.~M. 1998, \aap, 340, 300

\bibitem[{Roberge {et~al.}(2002)Roberge, {Feldman}, {Lecavelier des Etangs},
  {Vidal-Madjar}, {Deleuil}, {Bouret}, {Ferlet}, \& {Moos}}]{roberge2002}
Roberge, A., {Feldman}, P.~D., {Lecavelier des Etangs}, A., {et~al.} 2002,
  \apj, 568, 343

\bibitem[{{Rodgers} {et~al.}(2002){Rodgers}, {Wooden}, {Grinin}, {Shakhovsky},
  \& {Natta}}]{rodgers2002}
{Rodgers}, B., {Wooden}, D.~H., {Grinin}, V., {Shakhovsky}, D., \& {Natta}, A.
  2002, \apj, 564, 405

\bibitem[{Rostopchina {et~al.}(2000)Rostopchina, Grinin, Shakhovskoy, The, \&
  Minikulov}]{rostopchina2000}
Rostopchina, A.~N., Grinin, V.~P., Shakhovskoy, D.~N., The, P.~S., \&
  Minikulov, N.~K. 2000, Astronomy Reports, 44, 365

\bibitem[{{Shu} {et~al.}(2000){Shu}, {Najita}, {Shang}, \& {Li}}]{shu2000}
{Shu}, F.~H., {Najita}, J.~R., {Shang}, H., \& {Li}, Z.-Y. 2000, Protostars and
  Planets IV, 789

\bibitem[{{Spitzer}(1978)}]{spitzer1978}
{Spitzer}, L. 1978, {Physical processes in the interstellar medium} (New York
  Wiley-Interscience, 1978.)

\bibitem[{{van den Ancker} {et~al.}(2001){van den Ancker}, {Meeus}, {Cami},
  {Waters}, \& {Waelkens}}]{vandenancker2001}
{van den Ancker}, M.~E., {Meeus}, G., {Cami}, J., {Waters}, L.~B.~F.~M., \&
  {Waelkens}, C. 2001, \aap, 369, L17

\end{thebibliography}


\Online



\begin{table*}
\caption{
Identified Transient Absorption Components (TACs) in BF~Ori.
Column 1 gives the corresponding Balmer or metallic line, Col. 2  gives the Julian Date ($-$2\,450\,000), Col. 3 represents the event assigned to the particular absorptions (see text Sect.~\ref{circumstellar_contribution}), Cols. 4 to 6 give the parameters of each transient absorption as estimated from the multigaussian fit of the normalized residual absorption $R$: $v$, radial velocity, \deltav, FWHM, and $R_{\rm max}$, the strength of the absorption (peak of the gaussian).
``0'' in Col. 3 corresponds to the narrow IS absorptions, while ``--'' means that the absorption is not associated with a particular event.}
\label{master_table_bfori}
\centerline{
\tiny
\begin{tabular}{lllrrl}
\hline
\hline
Line & JD & Event & $v$ (\kms) & \deltav\ (\kms) & $R_{\rm max}$ \\
\hline
H$\beta$             & 1112.6324 & 1 &  -8 & 184 & 0.95 \\
H$\gamma$            & 1112.6324 & 1 &   7 & 183 & 0.95 \\
H$\delta$            & 1112.6324 & 1 &   4 & 157 & 0.93 \\
H$\zeta$             & 1112.6324 & 1 &  11 & 120 & 0.79 \\
\ion{Na}{i} D2       & 1112.6324 & 1 &  -1 &  66 & 0.83 \\
\ion{Na}{i} D1       & 1112.6324 & 1 &  -2 &  60 & 0.75 \\
\ion{Fe}{ii} 4924\AA & 1112.6324 & 1 &   3 &  83 & 0.64 \\
\ion{Fe}{ii} 5018\AA & 1112.6324 & 1 &   5 &  92 & 0.71 \\
\ion{Fe}{ii} 5169\AA & 1112.6324 & 1 &   2 &  93 & 0.76 \\
\ion{Ti}{ii} 4444\AA & 1112.6324 & 1 &  -2 &  61 & 0.33 \\
\ion{Ti}{ii} 4572\AA & 1112.6324 & 1 &  -3 &  58 & 0.35 \\
\ion{Sc}{ii} 4247\AA & 1112.6324 & 1 &  -1 &  66 & 0.30 \\
\ion{Ca}{i} 4227\AA  & 1112.6324 & 1 &   3 &  62 & 0.23 \\
H$\epsilon$          & 1112.6324 & --&  30 & 132 & 0.75 \\
\ion{Ca}{ii} K       & 1112.6324 & --& -52 &  70 & 0.54 \\
\ion{Ca}{ii} K       & 1112.6324 & --&  92 & 172 & 0.61 \\
\ion{Ca}{ii} H       & 1112.6324 & --& -40 &  53 & 0.71 \\
H$\beta$             & 1113.6515 & 1 &  27 & 194 & 0.86 \\
H$\gamma$            & 1113.6515 & 1 &  33 & 189 & 0.86 \\
H$\delta$            & 1113.6515 & 1 &  33 & 195 & 0.79 \\
H$\zeta$             & 1113.6515 & 1 &  39 & 139 & 0.65 \\
\ion{Na}{i} D2       & 1113.6515 & 1 &   6 &  57 & 0.80 \\
\ion{Na}{i} D1       & 1113.6515 & 1 &   5 &  51 & 0.69 \\
\ion{Fe}{ii} 4924\AA & 1113.6515 & 1 &  15 &  98 & 0.54 \\
\ion{Fe}{ii} 5018\AA & 1113.6515 & 1 &  20 & 111 & 0.62 \\
\ion{Fe}{ii} 5169\AA & 1113.6515 & 1 &  19 & 116 & 0.66 \\
\ion{Ti}{ii} 4444\AA & 1113.6515 & 1 &  17 &  76 & 0.24 \\
\ion{Ti}{ii} 4572\AA & 1113.6515 & 1 &  13 &  70 & 0.24 \\
\ion{Sc}{ii} 4247\AA & 1113.6515 & 1 &  16 &  83 & 0.23 \\
\ion{Ca}{i} 4227\AA  & 1113.6515 & 1 &  13 &  51 & 0.15 \\
H$\epsilon$          & 1113.6515 & --&  73 & 172 & 0.63 \\
\ion{Ca}{ii} K       & 1113.6515 & --& -74 & 134 & 0.28 \\
\ion{Ca}{ii} K       & 1113.6515 & --&  96 & 167 & 0.59 \\
H$\beta$             & 1209.5542 & 2 &  40 & 218 & 0.83 \\
H$\gamma$            & 1209.5542 & 2 &  40 & 229 & 0.88 \\
H$\delta$            & 1209.5542 & 2 &  40 & 217 & 0.82 \\
H$\epsilon$          & 1209.5542 & 2 &  58 & 176 & 0.60 \\
H$\zeta$             & 1209.5542 & 2 &  58 & 184 & 0.60 \\
\ion{Na}{i} D2       & 1209.5542 & 2 &  55 & 152 & 0.21 \\
\ion{Na}{i} D1       & 1209.5542 & 2 &  38 & 135 & 0.18 \\
\ion{Fe}{ii} 4924\AA & 1209.5542 & 2 &  46 & 188 & 0.36 \\
\ion{Fe}{ii} 5018\AA & 1209.5542 & 2 &  54 & 196 & 0.44 \\
\hline
\end{tabular}
\hspace{1.5cm}
\begin{tabular}{lllrrl}
\hline
\hline
Line & JD & Event & $v$ (\kms) & \deltav\ (\kms) & $R_{\rm max}$ \\
\hline
\ion{Fe}{ii} 5169\AA & 1209.5542 & 2 &  50 & 199 & 0.48 \\
\ion{Ti}{ii} 4444\AA & 1209.5542 & 2 &  38 & 185 & 0.14 \\
\ion{Ti}{ii} 4572\AA & 1209.5542 & 2 &  52 & 189 & 0.13 \\
\ion{Sc}{ii} 4247\AA & 1209.5542 & 2 &  21 &  75 & 0.22 \\
\ion{Na}{i} D2       & 1209.5542 & 0 &   1 &  31 & 0.58 \\
\ion{Na}{i} D1       & 1209.5542 & 0 &   1 &  22 & 0.51 \\
\ion{Fe}{ii} 4924\AA & 1209.5542 & 0 &   4 &  20 & 0.09 \\
\ion{Fe}{ii} 5018\AA & 1209.5542 & 0 &   4 &  18 & 0.11 \\
\ion{Fe}{ii} 5169\AA & 1209.5542 & 0 &   3 &  24 & 0.12 \\
\ion{Ca}{ii} K       & 1209.5542 & --& -95 & 136 & 0.33 \\
\ion{Ca}{ii} K       & 1209.5542 & --&  96 & 172 & 0.71 \\
H$\beta$             & 1210.4571 & 2 &  38 & 155 & 0.95 \\
H$\gamma$            & 1210.4571 & 2 &  25 & 185 & 0.97 \\
H$\delta$            & 1210.4571 & 2 &  29 & 155 & 0.95 \\
H$\epsilon$          & 1210.4571 & 2 &  32 & 130 & 0.79 \\
H$\zeta$             & 1210.4571 & 2 &  23 & 145 & 0.79 \\
\ion{Na}{i} D2       & 1210.4571 & 2 &  13 &  83 & 0.47 \\
\ion{Na}{i} D1       & 1210.4571 & 2 &  16 &  78 & 0.32 \\
\ion{Fe}{ii} 4924\AA & 1210.4571 & 2 &  23 & 110 & 0.52 \\
\ion{Fe}{ii} 5018\AA & 1210.4571 & 2 &  23 & 115 & 0.59 \\
\ion{Fe}{ii} 5169\AA & 1210.4571 & 2 &  22 & 118 & 0.65 \\
\ion{Ti}{ii} 4444\AA & 1210.4571 & 2 &  20 &  87 & 0.22 \\
\ion{Ti}{ii} 4572\AA & 1210.4571 & 2 &  12 & 117 & 0.21 \\
\ion{Sc}{ii} 4247\AA & 1210.4571 & 2 &  14 &  98 & 0.21 \\
H$\beta$             & 1210.4571 & 3 & -52 &  62 & 0.32 \\
H$\gamma$            & 1210.4571 & 3 & -67 &  94 & 0.19 \\
H$\delta$            & 1210.4571 & 3 & -61 &  90 & 0.27 \\
H$\epsilon$          & 1210.4571 & 3 & -65 &  43 & 0.15 \\
H$\zeta$             & 1210.4571 & 3 & -89 &  63 & 0.15 \\
\ion{Ca}{ii} K       & 1210.4571 & 3 & -77 &  92 & 0.47 \\
\ion{Ca}{ii} H       & 1210.4571 & 3 & -67 &  93 & 0.47 \\
\ion{Fe}{ii} 4924\AA & 1210.4571 & 3 & -57 &  77 & 0.08 \\
\ion{Fe}{ii} 5018\AA & 1210.4571 & 3 & -53 &  55 & 0.07 \\
\ion{Fe}{ii} 5169\AA & 1210.4571 & 3 & -54 &  48 & 0.08 \\
\ion{Ti}{ii} 4444\AA & 1210.4571 & 3 & -42 &  46 & 0.09 \\
\ion{Ti}{ii} 4572\AA & 1210.4571 & 3 & -34 &  25 & 0.04 \\
\ion{Na}{i} D2       & 1210.4571 & 0 &  -1 &  15 & 0.41 \\
\ion{Na}{i} D1       & 1210.4571 & 0 &   0 &  14 & 0.47 \\
\ion{Fe}{ii} 4924\AA & 1210.4571 & 0 &   3 &  10 & 0.05 \\
\ion{Fe}{ii} 5018\AA & 1210.4571 & 0 &   2 &  13 & 0.06 \\
\ion{Fe}{ii} 5169\AA & 1210.4571 & 0 &   2 &  13 & 0.06 \\
\ion{Ca}{ii} K       & 1210.4571 & --&  78 & 146 & 0.69 \\
\hline
\end{tabular}}
\end{table*}


\begin{table*}
\caption{
Identified Transient Absorption Components (TACs) in SV~Cep. 
Details as for Table~\ref{master_table_bfori}.}
\label{master_table_svcep}
\centerline{
\tiny
\begin{tabular}{lllrrl}
\hline
\hline
Line & JD & Event & $v$ (\kms) & \deltav\ (\kms) & $R_{\rm max}$ \\
\hline
H$\beta$             &  950.6668 & 1 &  35 &  52 & 0.58 \\
H$\gamma$            &  950.6668 & 1 &  40 &  45 & 0.35 \\
H$\delta$            &  950.6668 & 1 &  40 &  46 & 0.33 \\
H$\epsilon$          &  950.6668 & 1 &  31 &  50 & 0.38 \\
H$\zeta$             &  950.6668 & 1 &  40 &  80 & 0.28 \\
\ion{Ca}{ii} K       &  950.6668 & 1 &  25 &  27 & 0.42 \\
\ion{Ca}{ii} H       &  950.6668 & 1 &  23 &  16 & 0.38 \\
\ion{Na}{i} D2       &  950.6668 & 1 &  26 &  14 & 0.25 \\
\ion{Fe}{ii} 4924\AA &  950.6668 & 1 &  23 &  25 & 0.15 \\
\ion{Fe}{ii} 5018\AA &  950.6668 & 1 &  25 &  27 & 0.22 \\
\ion{Fe}{ii} 5169\AA &  950.6668 & 1 &  24 &  28 & 0.25 \\
H$\beta$             &  950.6668 & 2 & -17 &  73 & 0.93 \\
H$\gamma$            &  950.6668 & 2 & -11 &  84 & 0.83 \\
H$\delta$            &  950.6668 & 2 & -13 &  79 & 0.67 \\
H$\epsilon$          &  950.6668 & 2 & -17 &  49 & 0.42 \\
H$\zeta$             &  950.6668 & 2 & -22 &  74 & 0.35 \\
\ion{Ca}{ii} K       &  950.6668 & 2 &  -8 &  28 & 0.73 \\
\ion{Ca}{ii} H       &  950.6668 & 2 &  -6 &  24 & 0.66 \\
\ion{Fe}{ii} 4924\AA &  950.6668 & 2 & -11 &  17 & 0.10 \\
\ion{Fe}{ii} 5018\AA &  950.6668 & 2 &  -9 &  20 & 0.15 \\
\ion{Fe}{ii} 5169\AA &  950.6668 & 2 & -10 &  16 & 0.20 \\
\ion{Na}{i} D2       &  950.6668 & 0 &   1 &  16 & 1.04 \\
\ion{Na}{i} D1       &  950.6668 & 0 &   1 &  13 & 1.05 \\
H$\beta$             & 1025.6260 & 3 &  99 & 276 & 0.34 \\
H$\gamma$            & 1025.6260 & 3 &  78 & 217 & 0.48 \\
H$\delta$            & 1025.6260 & 3 &  76 & 219 & 0.43 \\
H$\zeta$             & 1025.6260 & 3 &  77 & 210 & 0.31 \\
\ion{Ca}{ii} K       & 1025.6260 & 3 &  40 & 193 & 0.51 \\
\ion{Na}{i} D2       & 1025.6260 & 3 &  71 & 162 & 0.13 \\
\ion{Na}{i} D1       & 1025.6260 & 3 &  68 & 199 & 0.10 \\
\ion{Fe}{ii} 4924\AA & 1025.6260 & 3 &  51 & 197 & 0.09 \\
\ion{Fe}{ii} 5018\AA & 1025.6260 & 3 &  70 & 245 & 0.12 \\
\ion{Fe}{ii} 5169\AA & 1025.6260 & 3 &  60 & 207 & 0.12 \\
H$\beta$             & 1025.6260 & 4 &  -6 & 147 & 0.67 \\
H$\gamma$            & 1025.6260 & 4 & -21 & 139 & 0.54 \\
H$\delta$            & 1025.6260 & 4 & -16 & 133 & 0.48 \\
H$\epsilon$          & 1025.6260 & 4 &  15 & 231 & 0.67 \\
H$\zeta$             & 1025.6260 & 4 & -10 & 134 & 0.40 \\
\ion{Ca}{ii} K       & 1025.6260 & 4 &   0 &  29 & 0.41 \\
\ion{Ca}{ii} H       & 1025.6260 & 4 &  -2 &  19 & 0.66 \\
\ion{Fe}{ii} 4924\AA & 1025.6260 & 4 &  12 &  33 & 0.09 \\
\ion{Fe}{ii} 5018\AA & 1025.6260 & 4 &  12 &  39 & 0.14 \\
\ion{Fe}{ii} 5169\AA & 1025.6260 & 4 &  10 &  39 & 0.15 \\
\ion{Na}{i} D2       & 1025.6260 & 0 &   0 &  17 & 0.95 \\
\ion{Na}{i} D1       & 1025.6260 & 0 &  -0 &  14 & 0.96 \\
H$\beta$             & 1026.6684 & 3 &  80 & 202 & 0.75 \\
H$\gamma$            & 1026.6684 & 3 &  72 & 190 & 0.76 \\
H$\delta$            & 1026.6684 & 3 &  69 & 164 & 0.70 \\
H$\epsilon$          & 1026.6684 & 3 &  87 & 116 & 0.56 \\
H$\zeta$             & 1026.6684 & 3 &  66 & 152 & 0.55 \\
\ion{Ca}{ii} K       & 1026.6684 & 3 &  75 & 115 & 0.53 \\
\ion{Na}{i} D2       & 1026.6684 & 3 &  82 & 130 & 0.16 \\
\ion{Na}{i} D1       & 1026.6684 & 3 &  75 & 127 & 0.12 \\
\ion{Fe}{ii} 4924\AA & 1026.6684 & 3 &  84 &  87 & 0.11 \\
\ion{Fe}{ii} 5018\AA & 1026.6684 & 3 &  87 &  95 & 0.16 \\
\ion{Fe}{ii} 5169\AA & 1026.6684 & 3 &  86 &  83 & 0.17 \\
H$\beta$             & 1026.6684 & 4 &  -8 &  82 & 0.59 \\
H$\gamma$            & 1026.6684 & 4 & -11 &  68 & 0.42 \\
H$\delta$            & 1026.6684 & 4 & -13 &  60 & 0.35 \\
H$\epsilon$          & 1026.6684 & 4 & -18 & 112 & 0.59 \\
H$\zeta$             & 1026.6684 & 4 & -10 &  53 & 0.21 \\
\hline
\end{tabular}
\hspace{1.5cm}
\begin{tabular}{lllrrl}
\hline
\hline
Line & JD & Event & $v$ (\kms) & \deltav\ (\kms) & $R_{\rm max}$ \\
\hline
\ion{Ca}{ii} K       & 1026.6684 & 4 &  -6 &  46 & 0.49 \\
\ion{Ca}{ii} H       & 1026.6684 & 4 & -22 &  20 & 0.32 \\
\ion{Fe}{ii} 4924\AA & 1026.6684 & 4 &   1 &  40 & 0.13 \\
\ion{Fe}{ii} 5018\AA & 1026.6684 & 4 &  -0 &  43 & 0.19 \\
\ion{Fe}{ii} 5169\AA & 1026.6684 & 4 &   0 &  42 & 0.23 \\
\ion{Ca}{ii} K       & 1026.6684 & 0 &  -3 &  12 & 0.29 \\
\ion{Ca}{ii} H       & 1026.6684 & 0 &  -2 &  17 & 0.85 \\
\ion{Na}{i} D2       & 1026.6684 & 0 &   0 &  18 & 0.97 \\
\ion{Na}{i} D1       & 1026.6684 & 0 &  -1 &  14 & 0.99 \\
H$\beta$             & 1113.4730 & 5 & 138 &  38 & 0.04 \\
H$\gamma$            & 1113.4730 & 5 & 113 &  64 & 0.12 \\
H$\delta$            & 1113.4730 & 5 & 114 &  81 & 0.12 \\
H$\epsilon$          & 1113.4730 & 5 &  97 &  41 & 0.13 \\
H$\zeta$             & 1113.4730 & 5 &  85 &  20 & 0.13 \\
\ion{Ca}{ii} K       & 1113.4730 & 5 &  89 &  85 & 0.11 \\
\ion{Fe}{ii} 4924\AA & 1113.4730 & 5 &  87 &  63 & 0.04 \\
\ion{Fe}{ii} 5018\AA & 1113.4730 & 5 &  90 & 292 & 0.05 \\
\ion{Fe}{ii} 5169\AA & 1113.4730 & 5 &  67 & 173 & 0.04 \\
H$\beta$             & 1113.4730 & 6 & -28 & 125 & 0.99 \\
H$\gamma$            & 1113.4730 & 6 & -21 & 117 & 0.94 \\
H$\delta$            & 1113.4730 & 6 & -22 & 106 & 0.86 \\
H$\epsilon$          & 1113.4730 & 6 & -13 &  85 & 0.65 \\
H$\zeta$             & 1113.4730 & 6 & -18 &  95 & 0.52 \\
\ion{Ca}{ii} K       & 1113.4730 & 6 & -19 &  77 & 0.56 \\
\ion{Ca}{ii} H       & 1113.4730 & 6 & -11 &  65 & 0.53 \\
\ion{Na}{i} D2       & 1113.4730 & 6 & -26 &  12 & 0.14 \\
\ion{Na}{i} D1       & 1113.4730 & 6 & -27 &  30 & 0.07 \\
\ion{Fe}{ii} 4924\AA & 1113.4730 & 6 & -15 &  78 & 0.15 \\
\ion{Fe}{ii} 5018\AA & 1113.4730 & 6 & -14 &  70 & 0.17 \\
\ion{Fe}{ii} 5169\AA & 1113.4730 & 6 & -23 &  77 & 0.18 \\
\ion{Ca}{ii} K       & 1113.4730 & 0 &  -3 &  15 & 0.44 \\
\ion{Ca}{ii} H       & 1113.4730 & 0 &  -1 &  12 & 0.50 \\
\ion{Na}{i} D2       & 1113.4730 & 0 &  -1 &  16 & 1.04 \\
\ion{Na}{i} D1       & 1113.4730 & 0 &  -1 &  14 & 1.03 \\
\ion{Fe}{ii} 4924\AA & 1113.4730 & 0 &   2 &  13 & 0.09 \\
\ion{Fe}{ii} 5018\AA & 1113.4730 & 0 &  -1 &  14 & 0.11 \\
\ion{Fe}{ii} 5169\AA & 1113.4730 & 0 &  -2 &  23 & 0.16 \\
H$\beta$             & 1209.3372 & 7 & 198 & 349 & 0.33 \\
H$\gamma$            & 1209.3372 & 7 & 164 & 315 & 0.32 \\
H$\delta$            & 1209.3372 & 7 & 134 & 334 & 0.30 \\
H$\epsilon$          & 1209.3372 & 7 & 141 & 298 & 0.29 \\
H$\zeta$             & 1209.3372 & 7 & 143 & 313 & 0.24 \\
\ion{Ca}{ii} K       & 1209.3372 & 7 & 172 & 269 & 0.19 \\
\ion{Fe}{ii} 4924\AA & 1209.3372 & 7 & 136 & 324 & 0.05 \\
\ion{Fe}{ii} 5018\AA & 1209.3372 & 7 & 156 & 253 & 0.07 \\
\ion{Fe}{ii} 5169\AA & 1209.3372 & 7 & 167 & 278 & 0.06 \\
H$\beta$             & 1209.3372 & 8 &  -4 &  99 & 0.97 \\
H$\gamma$            & 1209.3372 & 8 &   3 & 111 & 0.77 \\
H$\delta$            & 1209.3372 & 8 &   6 & 110 & 0.60 \\
H$\epsilon$          & 1209.3372 & 8 &   6 & 100 & 0.51 \\
H$\zeta$             & 1209.3372 & 8 &   9 &  99 & 0.46 \\
\ion{Ca}{ii} K       & 1209.3372 & 8 &  12 & 115 & 0.40 \\
\ion{Na}{i} D2       & 1209.3372 & 8 &  40 &  96 & 0.07 \\
\ion{Na}{i} D1       & 1209.3372 & 8 &  33 &  51 & 0.06 \\
\ion{Fe}{ii} 4924\AA & 1209.3372 & 8 &  19 &  76 & 0.09 \\
\ion{Fe}{ii} 5018\AA & 1209.3372 & 8 &  16 &  78 & 0.13 \\
\ion{Fe}{ii} 5169\AA & 1209.3372 & 8 &  17 &  74 & 0.15 \\
\ion{Ca}{ii} K       & 1209.3372 & 0 &  -1 &  18 & 0.43 \\
\ion{Ca}{ii} H       & 1209.3372 & 0 &   1 &  22 & 0.77 \\
\ion{Na}{i} D2       & 1209.3372 & 0 &   1 &  20 & 0.87 \\
\ion{Na}{i} D1       & 1209.3372 & 0 &   1 &  15 & 0.93 \\
\hline
\end{tabular}}
\end{table*}


\begin{table*}
\caption{
Identified Transient Absorption Components (TACs) in WW~Vul.
Details as for Table~\ref{master_table_bfori}.}
\label{master_table_wwvul}
\centerline{
\tiny
\begin{tabular}{lllrrl}
\hline
\hline
Line & JD & Event & $v$ (\kms) & \deltav\ (\kms)  & $R_{\rm max}$ \\
\hline
H$\beta$             &  950.6176 & 1 &  131 &  95 & 0.10 \\
H$\gamma$            &  950.6176 & 1 &   71 & 162 & 0.41 \\
H$\delta$            &  950.6176 & 1 &   58 & 174 & 0.44 \\
H$\epsilon$          &  950.6176 & 1 &   59 & 164 & 0.40 \\
H$\zeta$             &  950.6176 & 1 &   69 & 168 & 0.38 \\
\ion{Ca}{ii} K       &  950.6176 & 1 &   74 & 170 & 0.34 \\
\ion{Fe}{ii} 4924\AA &  950.6176 & 1 &   40 & 186 & 0.16 \\
\ion{Fe}{ii} 5018\AA &  950.6176 & 1 &   45 & 203 & 0.20 \\
\ion{Fe}{ii} 5169\AA &  950.6176 & 1 &   44 & 207 & 0.19 \\
\ion{Fe}{ii} 4924\AA &  950.6176 & 2 &  -13 &  36 & 0.17 \\
\ion{Fe}{ii} 5018\AA &  950.6176 & 2 &  -12 &  41 & 0.22 \\
\ion{Fe}{ii} 5169\AA &  950.6176 & 2 &  -12 &  41 & 0.26 \\
H$\beta$             &  950.6176 & 3 &  -25 & 110 & 1.27 \\
H$\gamma$            &  950.6176 & 3 &  -34 & 113 & 0.93 \\
H$\delta$            &  950.6176 & 3 &  -34 & 106 & 0.73 \\
H$\epsilon$          &  950.6176 & 3 &  -30 &  87 & 0.61 \\
H$\zeta$             &  950.6176 & 3 &  -31 & 104 & 0.62 \\
\ion{Ca}{ii} K       &  950.6176 & 3 &  -28 &  89 & 0.61 \\
\ion{Ca}{ii} H       &  950.6176 & 3 &  -11 &  58 & 0.98 \\
\ion{Na}{i} D2       &  950.6176 & 3 &  -34 &  91 & 0.16 \\
\ion{Na}{i} D1       &  950.6176 & 3 &  -22 &  59 & 0.12 \\
H$\beta$             &  950.6176 & 4 & -104 &  58 & 0.63 \\
H$\gamma$            &  950.6176 & 4 &  -97 &  55 & 0.43 \\
H$\delta$            &  950.6176 & 4 &  -90 &  55 & 0.39 \\
H$\zeta$             &  950.6176 & 4 &  -82 &  49 & 0.25 \\
\ion{Ca}{ii} K       &  950.6176 & 4 &  -86 &  42 & 0.31 \\
\ion{Ca}{ii} H       &  950.6176 & 4 &  -74 &  55 & 0.63 \\
\ion{Fe}{ii} 4924\AA &  950.6176 & 4 &  -67 &  45 & 0.09 \\
\ion{Fe}{ii} 5018\AA &  950.6176 & 4 &  -68 &  46 & 0.12 \\
\ion{Fe}{ii} 5169\AA &  950.6176 & 4 &  -66 &  41 & 0.13 \\
\ion{Ca}{ii} K       &  950.6176 & 0 &   -4 &  23 & 0.21 \\
\ion{Ca}{ii} H       &  950.6176 & 0 &   -1 &   7 & 0.09 \\
\ion{Na}{i} D2       &  950.6176 & 0 &   -2 &  22 & 0.86 \\
\ion{Na}{i} D1       &  950.6176 & 0 &   -1 &  16 & 0.80 \\
H$\beta$             &  951.6232 & 1 &   93 &  75 & 0.33 \\
H$\gamma$            &  951.6232 & 1 &   72 & 109 & 0.53 \\
H$\delta$            &  951.6232 & 1 &   42 & 136 & 0.68 \\
H$\epsilon$          &  951.6232 & 1 &   70 &  91 & 0.34 \\
\ion{Ca}{ii} K       &  951.6232 & 1 &   30 & 142 & 0.57 \\
\ion{Na}{i} D2       &  951.6232 & 1 &   59 &  62 & 0.14 \\
\ion{Na}{i} D1       &  951.6232 & 1 &   67 &  43 & 0.12 \\
\ion{Fe}{ii} 4924\AA &  951.6232 & 1 &   45 & 112 & 0.18 \\
\ion{Fe}{ii} 5018\AA &  951.6232 & 1 &   44 & 111 & 0.26 \\
\ion{Fe}{ii} 5169\AA &  951.6232 & 1 &   46 & 113 & 0.27 \\
\ion{Ca}{ii} K       &  951.6232 & 2 &  -21 &  53 & 0.36 \\
\ion{Ca}{ii} H       &  951.6232 & 2 &  -12 &  46 & 0.74 \\
\ion{Na}{i} D2       &  951.6232 & 2 &  -14 &  59 & 0.50 \\
\ion{Na}{i} D1       &  951.6232 & 2 &  -10 &  58 & 0.36 \\
\ion{Fe}{ii} 4924\AA &  951.6232 & 2 &  -20 &  42 & 0.12 \\
\ion{Fe}{ii} 5018\AA &  951.6232 & 2 &  -20 &  52 & 0.21 \\
\ion{Fe}{ii} 5169\AA &  951.6232 & 2 &  -23 &  66 & 0.33 \\
H$\beta$             &  951.6232 & 3 &  -30 & 134 & 1.10 \\
H$\gamma$            &  951.6232 & 3 &  -39 & 136 & 0.96 \\
H$\delta$            &  951.6232 & 3 &  -53 & 107 & 0.70 \\
H$\epsilon$          &  951.6232 & 3 &  -21 & 129 & 0.79 \\
H$\zeta$             &  951.6232 & 3 &  -49 &  52 & 0.21 \\
\ion{Ca}{ii} K       &  951.6232 & 3 &  -64 &  41 & 0.31 \\
\ion{Ca}{ii} H       &  951.6232 & 3 &  -57 &  50 & 0.57 \\
\ion{Fe}{ii} 4924\AA &  951.6232 & 3 &  -40 & 135 & 0.13 \\
\ion{Fe}{ii} 5018\AA &  951.6232 & 3 &  -63 & 127 & 0.12 \\
\ion{Fe}{ii} 5169\AA &  951.6232 & 3 &  -91 &  84 & 0.10 \\
H$\beta$             &  951.6232 & 4 & -174 &  63 & 0.14 \\
H$\gamma$            &  951.6232 & 4 & -170 &  81 & 0.19 \\
H$\delta$            &  951.6232 & 4 & -156 &  78 & 0.18 \\
H$\zeta$             &  951.6232 & 4 & -137 & 119 & 0.10 \\
\ion{Ca}{ii} K       &  951.6232 & 4 & -122 & 113 & 0.17 \\
\ion{Ca}{ii} H       &  951.6232 & 4 & -139 &  70 & 0.10 \\
\ion{Ca}{ii} K       &  951.6232 & 0 &   -3 &  19 & 0.14 \\
\ion{Ca}{ii} H       &  951.6232 & 0 &   -4 &   9 & 0.07 \\
\ion{Na}{i} D2       &  951.6232 & 0 &   -0 &  15 & 0.54 \\
\ion{Na}{i} D1       &  951.6232 & 0 &    0 &  12 & 0.62 \\
\ion{Fe}{ii} 4924\AA &  951.6232 & 0 &   11 &  11 & 0.04 \\
\ion{Fe}{ii} 5018\AA &  951.6232 & 0 &   11 &   8 & 0.06 \\
\ion{Fe}{ii} 5169\AA &  951.6232 & 0 &    8 &   8 & 0.03 \\
H$\zeta$             &  951.6232 & --&   10 & 163 & 0.75 \\
\ion{Ca}{ii} K       & 1023.5186 & 5 &   41 & 106 & 0.46 \\
\ion{Ca}{ii} H       & 1023.5186 & 5 &   28 &  28 & 0.28 \\
\ion{Fe}{ii} 4924\AA & 1023.5186 & 5 &   37 & 115 & 0.08 \\
\ion{Fe}{ii} 5018\AA & 1023.5186 & 5 &   50 & 132 & 0.11 \\
\ion{Fe}{ii} 5169\AA & 1023.5186 & 5 &   40 & 127 & 0.11 \\
\ion{Ti}{ii} 4444\AA & 1023.5186 & 5 &   63 & 113 & 0.02 \\
H$\beta$             & 1023.5186 & 6 &  -11 & 157 & 1.31 \\
H$\gamma$            & 1023.5186 & 6 &   -9 & 155 & 1.09 \\
H$\delta$            & 1023.5186 & 6 &   -9 & 131 & 0.99 \\
\hline
\end{tabular}
\hspace{1.5cm}
\begin{tabular}{lllrrl}
\hline
\hline
Line & JD & Event & $v$ (\kms) & \deltav\ (\kms)  & $R_{\rm max}$ \\
\hline
H$\epsilon$          & 1023.5186 & 6 &   -3 &  99 & 0.98 \\
H$\zeta$             & 1023.5186 & 6 &   -9 & 101 & 0.85 \\
\ion{Ca}{ii} K       & 1023.5186 & 6 &  -11 &  62 & 0.66 \\
\ion{Ca}{ii} H       & 1023.5186 & 6 &   -6 &  46 & 1.15 \\
\ion{Fe}{ii} 4924\AA & 1023.5186 & 6 &   -6 &  34 & 0.32 \\
\ion{Fe}{ii} 5018\AA & 1023.5186 & 6 &   -6 &  40 & 0.41 \\
\ion{Fe}{ii} 5169\AA & 1023.5186 & 6 &   -5 &  39 & 0.45 \\
\ion{Ti}{ii} 4444\AA & 1023.5186 & 6 &   -6 &  36 & 0.05 \\
\ion{Ti}{ii} 4572\AA & 1023.5186 & 6 &   -7 &  37 & 0.05 \\
\ion{Ca}{ii} K       & 1023.5186 & 7 &  -46 &  28 & 0.41 \\
\ion{Ca}{ii} H       & 1023.5186 & 7 &  -41 &  31 & 0.85 \\
\ion{Na}{i} D2       & 1023.5186 & 7 &  -30 &  25 & 0.50 \\
\ion{Na}{i} D1       & 1023.5186 & 7 &  -23 &  34 & 0.35 \\
\ion{Fe}{ii} 4924\AA & 1023.5186 & 7 &  -35 &  24 & 0.30 \\
\ion{Fe}{ii} 5018\AA & 1023.5186 & 7 &  -35 &  23 & 0.33 \\
\ion{Fe}{ii} 5169\AA & 1023.5186 & 7 &  -35 &  24 & 0.40 \\
\ion{Ti}{ii} 4444\AA & 1023.5186 & 7 &  -34 &  21 & 0.04 \\
\ion{Ti}{ii} 4572\AA & 1023.5186 & 7 &  -34 &  22 & 0.04 \\
\ion{Na}{i} D2       & 1023.5186 & 0 &   -1 &  22 & 0.93 \\
\ion{Na}{i} D1       & 1023.5186 & 0 &    0 &  14 & 0.86 \\
\ion{Fe}{ii} 4924\AA & 1112.3689 & 8 &   86 & 162 & 0.09 \\
\ion{Fe}{ii} 5018\AA & 1112.3689 & 8 &   96 & 195 & 0.13 \\
\ion{Fe}{ii} 5169\AA & 1112.3689 & 8 &  113 & 248 & 0.11 \\
\ion{Ti}{ii} 4444\AA & 1112.3689 & 8 &   91 & 163 & 0.03 \\
\ion{Ti}{ii} 4572\AA & 1112.3689 & 8 &   78 & 224 & 0.03 \\
H$\beta$             & 1112.3689 & 9 &    7 & 156 & 0.98 \\
H$\gamma$            & 1112.3689 & 9 &   17 & 159 & 0.97 \\
H$\delta$            & 1112.3689 & 9 &   17 & 152 & 0.99 \\
H$\epsilon$          & 1112.3689 & 9 &   11 & 174 & 1.04 \\
H$\zeta$             & 1112.3689 & 9 &   24 & 145 & 0.93 \\
\ion{Ca}{ii} K       & 1112.3689 & 9 &   23 & 148 & 0.78 \\
\ion{Na}{i} D2       & 1112.3689 & 9 &   21 &  83 & 0.59 \\
\ion{Na}{i} D1       & 1112.3689 & 9 &   22 &  72 & 0.45 \\
\ion{Fe}{ii} 4924\AA & 1112.3689 & 9 &   18 &  77 & 0.32 \\
\ion{Fe}{ii} 5018\AA & 1112.3689 & 9 &   17 &  82 & 0.39 \\
\ion{Fe}{ii} 5169\AA & 1112.3689 & 9 &   18 &  85 & 0.45 \\
\ion{Fe}{i} 4046\AA  & 1112.3689 & 9 &   32 &  76 & 0.03 \\
\ion{Ti}{ii} 4444\AA & 1112.3689 & 9 &   14 &  57 & 0.08 \\
\ion{Ti}{ii} 4572\AA & 1112.3689 & 9 &   16 &  62 & 0.09 \\
\ion{Sc}{ii} 4247\AA & 1112.3689 & 9 &   35 &  79 & 0.08 \\
\ion{Ca}{i} 4227\AA  & 1112.3689 & 9 &   17 &  66 & 0.03 \\
\ion{Ca}{ii} K       & 1112.3689 & 0 &   -7 &  25 & 0.16 \\
\ion{Ca}{ii} H       & 1112.3689 & 0 &   -5 &  56 & 0.89 \\
\ion{Na}{i} D2       & 1112.3689 & 0 &   -2 &  14 & 0.51 \\
\ion{Na}{i} D1       & 1112.3689 & 0 &   -1 &  13 & 0.62 \\
\ion{Fe}{ii} 4924\AA & 1112.3689 & 0 &    0 &  13 & 0.11 \\
\ion{Fe}{ii} 5018\AA & 1112.3689 & 0 &    1 &  12 & 0.12 \\
\ion{Fe}{ii} 5169\AA & 1112.3689 & 0 &   -0 &  13 & 0.15 \\
\ion{Sc}{ii} 4247\AA & 1112.3689 & 0 &    2 &  31 & 0.07 \\
H$\gamma$            & 1113.3958 & 8 &  129 & 248 & 0.42 \\
H$\delta$            & 1113.3958 & 8 &  106 & 233 & 0.53 \\
H$\epsilon$          & 1113.3958 & 8 &   85 & 257 & 0.62 \\
H$\zeta$             & 1113.3958 & 8 &  106 & 220 & 0.47 \\
\ion{Ca}{ii} K       & 1113.3958 & 8 &  108 & 239 & 0.46 \\
\ion{Na}{i} D2       & 1113.3958 & 8 &   82 & 187 & 0.14 \\
\ion{Na}{i} D1       & 1113.3958 & 8 &  129 & 185 & 0.11 \\
\ion{Fe}{ii} 4924\AA & 1113.3958 & 8 &   80 & 236 & 0.18 \\
\ion{Fe}{ii} 5018\AA & 1113.3958 & 8 &   90 & 263 & 0.24 \\
\ion{Fe}{ii} 5169\AA & 1113.3958 & 8 &  116 & 291 & 0.24 \\
\ion{Fe}{i} 4046\AA  & 1113.3958 & 8 &  127 & 116 & 0.04 \\
\ion{Ti}{ii} 4444\AA & 1113.3958 & 8 &   74 & 231 & 0.06 \\
\ion{Ti}{ii} 4572\AA & 1113.3958 & 8 &  107 & 253 & 0.06 \\
\ion{Sc}{ii} 4247\AA & 1113.3958 & 8 &  123 & 140 & 0.06 \\
\ion{Ca}{i} 4227\AA  & 1113.3958 & 8 &  134 & 109 & 0.03 \\
H$\beta$             & 1113.3958 & 9 &   -4 & 124 & 0.96 \\
H$\gamma$            & 1113.3958 & 9 &   -4 & 121 & 0.73 \\
H$\delta$            & 1113.3958 & 9 &  -11 & 109 & 0.66 \\
H$\epsilon$          & 1113.3958 & 9 &    6 &  68 & 0.44 \\
H$\zeta$             & 1113.3958 & 9 &   -9 & 102 & 0.65 \\
\ion{Ca}{ii} K       & 1113.3958 & 9 &   -4 &  63 & 0.49 \\
\ion{Ca}{ii} H       & 1113.3958 & 9 &   -2 &  58 & 0.79 \\
\ion{Na}{i} D2       & 1113.3958 & 9 &    4 &  48 & 0.74 \\
\ion{Na}{i} D1       & 1113.3958 & 9 &    5 &  43 & 0.76 \\
\ion{Fe}{ii} 4924\AA & 1113.3958 & 9 &    4 &  43 & 0.34 \\
\ion{Fe}{ii} 5018\AA & 1113.3958 & 9 &    4 &  47 & 0.39 \\
\ion{Fe}{ii} 5169\AA & 1113.3958 & 9 &    4 &  49 & 0.46 \\
\ion{Fe}{i} 4046\AA  & 1113.3958 & 9 &    3 &  40 & 0.06 \\
\ion{Ti}{ii} 4444\AA & 1113.3958 & 9 &    6 &  36 & 0.09 \\
\ion{Ti}{ii} 4572\AA & 1113.3958 & 9 &    1 &  48 & 0.09 \\
\ion{Sc}{ii} 4247\AA & 1113.3958 & 9 &    6 &  55 & 0.12 \\
\ion{Ca}{i} 4227\AA  & 1113.3958 & 9 &    6 &  46 & 0.10 \\
H$\beta$             & 1113.3958 & --&  251 & 149 & 0.20 \\
\hline
\\
\end{tabular}}
\end{table*}


\begin{table*}
\caption{
Identified Transient Absorption Components (TACs) in XY~Per.
Details as for Table~\ref{master_table_bfori}.}
\label{master_table_xyper}
\centerline{
\tiny
\begin{tabular}{lllrrl}
\hline
\hline
Line & JD & Event & $v$ (\kms) & \deltav\ (\kms) & $R_{\rm max}$ \\
\hline
H$\beta$             & 1024.6728 & 2 &  -9 & 171 & 0.92 \\
H$\gamma$            & 1024.6728 & 2 &  -4 & 171 & 0.92 \\
H$\delta$            & 1024.6728 & 2 &  -3 & 163 & 0.88 \\
H$\epsilon$          & 1024.6728 & 2 &   2 & 147 & 0.72 \\
H$\zeta$             & 1024.6728 & 2 &  -5 & 146 & 0.70 \\
\ion{Ca}{ii} K       & 1024.6728 & 2 &  -7 & 142 & 0.77 \\
\ion{Ca}{ii} H       & 1024.6728 & 2 & -27 & 104 & 0.63 \\
\ion{Na}{i} D2       & 1024.6728 & 2 & -26 & 100 & 0.16 \\
\ion{Na}{i} D1       & 1024.6728 & 2 & -19 &  92 & 0.10 \\
\ion{Fe}{ii} 4924\AA & 1024.6728 & 2 & -15 & 118 & 0.25 \\
\ion{Fe}{ii} 5018\AA & 1024.6728 & 2 & -14 & 115 & 0.32 \\
\ion{Fe}{ii} 5169\AA & 1024.6728 & 2 & -16 & 114 & 0.35 \\
\ion{Ti}{ii} 4444\AA & 1024.6728 & 2 &   2 & 156 & 0.06 \\
\ion{Ti}{ii} 4572\AA & 1024.6728 & 2 & -11 & 144 & 0.06 \\
\ion{Sc}{ii} 4247\AA & 1024.6728 & 2 &  -3 & 133 & 0.04 \\
\ion{Ca}{ii} K       & 1024.6728 & 0 &  -0 &  18 & 0.23 \\
\ion{Ca}{ii} H       & 1024.6728 & 0 &  -1 &  25 & 0.44 \\
\ion{Na}{i} D2       & 1024.6728 & 0 &  -0 &  19 & 0.93 \\
\ion{Na}{i} D1       & 1024.6728 & 0 &   0 &  17 & 0.98 \\
\ion{Fe}{ii} 4924\AA & 1024.6728 & 0 &  -5 &  27 & 0.06 \\
\ion{Fe}{ii} 5018\AA & 1024.6728 & 0 &  -6 &  23 & 0.06 \\
\ion{Fe}{ii} 5169\AA & 1024.6728 & 0 &  -4 &  27 & 0.07 \\
\ion{Sc}{ii} 4247\AA & 1024.6728 & 0 &  -7 &   9 & 0.01 \\
H$\beta$             & 1025.6948 & 1 &  70 &  72 & 0.37 \\
H$\gamma$            & 1025.6948 & 1 &  74 &  58 & 0.16 \\
H$\delta$            & 1025.6948 & 1 &  66 &  93 & 0.21 \\
H$\epsilon$          & 1025.6948 & 1 &  60 &  59 & 0.23 \\
\ion{Fe}{ii} 4924\AA & 1025.6948 & 1 &  59 &  66 & 0.05 \\
\ion{Fe}{ii} 5018\AA & 1025.6948 & 1 &  73 &  66 & 0.05 \\
\ion{Fe}{ii} 5169\AA & 1025.6948 & 1 &  65 &  57 & 0.04 \\
\ion{Ti}{ii} 4444\AA & 1025.6948 & 1 &  34 & 113 & 0.06 \\
\ion{Ti}{ii} 4572\AA & 1025.6948 & 1 &  44 &  79 & 0.04 \\
\ion{Sc}{ii} 4247\AA & 1025.6948 & 1 &  37 &  88 & 0.04 \\
H$\beta$             & 1025.6948 & 2 & -10 &  90 & 0.90 \\
H$\gamma$            & 1025.6948 & 2 & -22 & 147 & 0.96 \\
H$\delta$            & 1025.6948 & 2 & -27 & 133 & 0.90 \\
H$\epsilon$          & 1025.6948 & 2 & -15 &  96 & 0.63 \\
H$\zeta$             & 1025.6948 & 2 & -11 & 141 & 0.70 \\
\ion{Ca}{ii} K       & 1025.6948 & 2 & -20 & 130 & 0.78 \\
\ion{Ca}{ii} H       & 1025.6948 & 2 & -20 &  94 & 0.73 \\
\ion{Na}{i} D2       & 1025.6948 & 2 & -28 & 104 & 0.24 \\
\ion{Na}{i} D1       & 1025.6948 & 2 & -29 &  89 & 0.16 \\
\ion{Fe}{ii} 4924\AA & 1025.6948 & 2 & -35 & 115 & 0.27 \\
\ion{Fe}{ii} 5018\AA & 1025.6948 & 2 & -29 & 117 & 0.34 \\
\ion{Fe}{ii} 5169\AA & 1025.6948 & 2 & -34 & 113 & 0.38 \\
\ion{Ti}{ii} 4444\AA & 1025.6948 & 2 & -44 &  84 & 0.05 \\
\ion{Ti}{ii} 4572\AA & 1025.6948 & 2 & -42 &  88 & 0.06 \\
\ion{Sc}{ii} 4247\AA & 1025.6948 & 2 & -42 &  69 & 0.03 \\
H$\beta$             & 1025.6948 & 3 & -88 &  82 & 0.65 \\
H$\gamma$            & 1025.6948 & 3 & -89 &  51 & 0.17 \\
H$\delta$            & 1025.6948 & 3 & -85 &  46 & 0.14 \\
H$\zeta$             & 1025.6948 & 3 & -68 &  58 & 0.14 \\
\ion{Ca}{ii} K       & 1025.6948 & 3 & -94 &  40 & 0.17 \\
\ion{Ca}{ii} H       & 1025.6948 & 3 & -85 &  51 & 0.28 \\
\ion{Fe}{ii} 5018\AA & 1025.6948 & 3 & -77 &  28 & 0.03 \\
\ion{Fe}{ii} 5169\AA & 1025.6948 & 3 & -79 &  29 & 0.04 \\
\ion{Ca}{ii} K       & 1025.6948 & 0 &   1 &  16 & 0.27 \\
\ion{Ca}{ii} H       & 1025.6948 & 0 &   1 &  14 & 0.32 \\
\ion{Na}{i} D2       & 1025.6948 & 0 &  -1 &  19 & 0.87 \\
\ion{Na}{i} D1       & 1025.6948 & 0 &  -1 &  18 & 0.96 \\
\ion{Fe}{ii} 4924\AA & 1025.6948 & 0 &   1 &  23 & 0.15 \\
\ion{Fe}{ii} 5018\AA & 1025.6948 & 0 &  -0 &  20 & 0.15 \\
\ion{Fe}{ii} 5169\AA & 1025.6948 & 0 &  -1 &  23 & 0.18 \\
\ion{Ti}{ii} 4444\AA & 1025.6948 & 0 &   1 &  20 & 0.04 \\
\ion{Ti}{ii} 4572\AA & 1025.6948 & 0 &   1 &  28 & 0.03 \\
\ion{Sc}{ii} 4247\AA & 1025.6948 & 0 &   2 &  22 & 0.02 \\
H$\beta$             & 1026.7065 & 1 &  69 &  75 & 0.24 \\
H$\gamma$            & 1026.7065 & 1 &  62 &  81 & 0.23 \\
H$\delta$            & 1026.7065 & 1 &  55 &  96 & 0.24 \\
H$\epsilon$          & 1026.7065 & 1 &  65 &  59 & 0.10 \\
H$\zeta$             & 1026.7065 & 1 &  51 &  72 & 0.20 \\
\ion{Ca}{ii} K       & 1026.7065 & 1 &  49 &  27 & 0.10 \\
\ion{Fe}{ii} 4924\AA & 1026.7065 & 1 &  30 &  76 & 0.10 \\
\ion{Fe}{ii} 5169\AA & 1026.7065 & 1 &  24 &  82 & 0.12 \\
\ion{Ti}{ii} 4444\AA & 1026.7065 & 1 &  32 & 113 & 0.04 \\
\ion{Ti}{ii} 4572\AA & 1026.7065 & 1 &  62 &  68 & 0.02 \\
\ion{Sc}{ii} 4247\AA & 1026.7065 & 1 &  58 &  51 & 0.02 \\
H$\beta$             & 1026.7065 & 2 & -58 & 124 & 0.79 \\
H$\gamma$            & 1026.7065 & 2 & -48 & 117 & 0.78 \\
H$\delta$            & 1026.7065 & 2 & -52 & 105 & 0.76 \\
H$\epsilon$          & 1026.7065 & 2 & -61 &  51 & 0.40 \\
H$\zeta$             & 1026.7065 & 2 & -51 &  84 & 0.57 \\
\hline
\end{tabular}
\hspace{1.5cm}
\begin{tabular}{lllrrl}
\hline
\hline
Line & JD & Event & $v$ (\kms) & \deltav\ (\kms) & $R_{\rm max}$ \\
\hline
\ion{Ca}{ii} K       & 1026.7065 & 2 & -66 &  91 & 0.55 \\
\ion{Ca}{ii} H       & 1026.7065 & 2 & -51 & 101 & 0.55 \\
\ion{Na}{i} D2       & 1026.7065 & 2 & -36 & 102 & 0.26 \\
\ion{Na}{i} D1       & 1026.7065 & 2 & -43 &  64 & 0.18 \\
\ion{Fe}{ii} 4924\AA & 1026.7065 & 2 & -52 &  83 & 0.37 \\
\ion{Fe}{ii} 5018\AA & 1026.7065 & 2 & -56 &  66 & 0.43 \\
\ion{Fe}{ii} 5169\AA & 1026.7065 & 2 & -54 &  82 & 0.45 \\
\ion{Ti}{ii} 4444\AA & 1026.7065 & 2 & -52 &  66 & 0.09 \\
\ion{Ti}{ii} 4572\AA & 1026.7065 & 2 & -53 &  68 & 0.10 \\
\ion{Sc}{ii} 4247\AA & 1026.7065 & 2 & -54 &  68 & 0.06 \\
H$\beta$             & 1026.7065 & 4 &  -3 &  54 & 0.42 \\
H$\gamma$            & 1026.7065 & 4 &   0 &  41 & 0.24 \\
H$\delta$            & 1026.7065 & 4 &   2 &  38 & 0.19 \\
H$\epsilon$          & 1026.7065 & 4 &  -4 &  38 & 0.21 \\
H$\zeta$             & 1026.7065 & 4 &   4 &  42 & 0.20 \\
\ion{Ca}{ii} K       & 1026.7065 & 4 &  -8 &  48 & 0.63 \\
\ion{Ca}{ii} H       & 1026.7065 & 4 &  -4 &  33 & 0.49 \\
\ion{Fe}{ii} 4924\AA & 1026.7065 & 4 &  -3 &  26 & 0.11 \\
\ion{Fe}{ii} 5018\AA & 1026.7065 & 4 &  -1 &  46 & 0.28 \\
\ion{Fe}{ii} 5169\AA & 1026.7065 & 4 &  -5 &  29 & 0.15 \\
\ion{Ti}{ii} 4444\AA & 1026.7065 & 4 &  -1 &  33 & 0.04 \\
\ion{Ti}{ii} 4572\AA & 1026.7065 & 4 &   5 &  50 & 0.05 \\
\ion{Sc}{ii} 4247\AA & 1026.7065 & 4 &  12 &  52 & 0.03 \\
\ion{Na}{i} D2       & 1026.7065 & 0 &  -0 &  17 & 0.89 \\
\ion{Na}{i} D1       & 1026.7065 & 0 &  -0 &  17 & 1.02 \\
H$\beta$             & 1112.4978 & 5 & 131 & 149 & 0.27 \\
H$\gamma$            & 1112.4978 & 5 &  83 & 206 & 0.41 \\
H$\delta$            & 1112.4978 & 5 & 118 & 173 & 0.30 \\
H$\zeta$             & 1112.4978 & 5 & 127 & 182 & 0.20 \\
\ion{Ca}{ii} K       & 1112.4978 & 5 & 119 & 177 & 0.29 \\
\ion{Fe}{ii} 4924\AA & 1112.4978 & 5 & 147 & 139 & 0.07 \\
\ion{Fe}{ii} 5018\AA & 1112.4978 & 5 & 123 & 190 & 0.13 \\
\ion{Fe}{ii} 5169\AA & 1112.4978 & 5 & 147 & 218 & 0.11 \\
\ion{Ti}{ii} 4444\AA & 1112.4978 & 5 & 122 & 209 & 0.02 \\
H$\beta$             & 1112.4978 & 6 &  30 & 129 & 0.77 \\
H$\gamma$            & 1112.4978 & 6 &  18 & 136 & 0.58 \\
H$\delta$            & 1112.4978 & 6 &  14 & 142 & 0.73 \\
H$\epsilon$          & 1112.4978 & 6 &  16 & 189 & 0.84 \\
H$\zeta$             & 1112.4978 & 6 &  17 & 139 & 0.59 \\
\ion{Ca}{ii} K       & 1112.4978 & 6 &  35 & 104 & 0.47 \\
\ion{Na}{i} D2       & 1112.4978 & 6 &  40 & 119 & 0.15 \\
\ion{Na}{i} D1       & 1112.4978 & 6 &  29 & 132 & 0.10 \\
\ion{Fe}{ii} 4924\AA & 1112.4978 & 6 &  26 & 124 & 0.25 \\
\ion{Fe}{ii} 5018\AA & 1112.4978 & 6 &  22 & 114 & 0.26 \\
\ion{Fe}{ii} 5169\AA & 1112.4978 & 6 &  18 & 121 & 0.28 \\
\ion{Fe}{i} 4046\AA  & 1112.4978 & 6 &  39 & 100 & 0.03 \\
\ion{Ti}{ii} 4444\AA & 1112.4978 & 6 &  32 & 117 & 0.06 \\
\ion{Ti}{ii} 4572\AA & 1112.4978 & 6 &  38 & 139 & 0.08 \\
\ion{Sc}{ii} 4247\AA & 1112.4978 & 6 &  36 & 137 & 0.07 \\
\ion{Ca}{i} 4227\AA  & 1112.4978 & 6 &  23 &  66 & 0.05 \\
H$\beta$             & 1112.4978 & 7 & -26 &  57 & 0.58 \\
H$\gamma$            & 1112.4978 & 7 & -24 &  44 & 0.34 \\
H$\delta$            & 1112.4978 & 7 & -21 &  37 & 0.23 \\
H$\zeta$             & 1112.4978 & 7 & -16 &  40 & 0.15 \\
\ion{Ca}{ii} K       & 1112.4978 & 7 & -34 &  33 & 0.53 \\
\ion{Ca}{ii} H       & 1112.4978 & 7 & -25 &  43 & 0.90 \\
\ion{Na}{i} D2       & 1112.4978 & 7 & -23 &  19 & 0.31 \\
\ion{Na}{i} D1       & 1112.4978 & 7 & -22 &  18 & 0.18 \\
\ion{Fe}{ii} 4924\AA & 1112.4978 & 7 & -22 &  24 & 0.18 \\
\ion{Fe}{ii} 5018\AA & 1112.4978 & 7 & -22 &  24 & 0.24 \\
\ion{Fe}{ii} 5169\AA & 1112.4978 & 7 & -22 &  22 & 0.27 \\
\ion{Ti}{ii} 4444\AA & 1112.4978 & 7 & -21 &  20 & 0.04 \\
\ion{Ti}{ii} 4572\AA & 1112.4978 & 7 & -24 &  23 & 0.04 \\
\ion{Ca}{ii} K       & 1112.4978 & 0 &  -2 &  30 & 0.48 \\
\ion{Ca}{ii} H       & 1112.4978 & 0 &   2 &  20 & 0.61 \\
\ion{Na}{i} D2       & 1112.4978 & 0 &   1 &  19 & 0.92 \\
\ion{Na}{i} D1       & 1112.4978 & 0 &   1 &  16 & 0.97 \\
H$\beta$             & 1113.5483 & 5 &  92 &  53 & 0.08 \\
H$\gamma$            & 1113.5483 & 5 &  76 & 107 & 0.26 \\
H$\delta$            & 1113.5483 & 5 &  75 & 123 & 0.28 \\
H$\zeta$             & 1113.5483 & 5 &  82 & 111 & 0.21 \\
\ion{Ca}{ii} K       & 1113.5483 & 5 &  76 &  51 & 0.11 \\
\ion{Fe}{ii} 4924\AA & 1113.5483 & 5 &  85 &  85 & 0.12 \\
\ion{Fe}{ii} 5018\AA & 1113.5483 & 5 &  79 & 136 & 0.14 \\
\ion{Fe}{ii} 5169\AA & 1113.5483 & 5 &  87 &  94 & 0.14 \\
\ion{Ti}{ii} 4444\AA & 1113.5483 & 5 &  82 & 115 & 0.03 \\
\ion{Ti}{ii} 4572\AA & 1113.5483 & 5 &  94 &  83 & 0.03 \\
H$\beta$             & 1113.5483 & 7 & -16 & 149 & 0.99 \\
H$\gamma$            & 1113.5483 & 7 & -23 & 133 & 0.93 \\
H$\delta$            & 1113.5483 & 7 & -25 & 126 & 0.89 \\
H$\epsilon$          & 1113.5483 & 7 &   1 & 133 & 0.51 \\
H$\zeta$             & 1113.5483 & 7 & -19 & 117 & 0.73 \\
\hline
\end{tabular}}
\end{table*}
\begin{table*}
\addtocounter{table}{-1}
\caption{(Cont.)}
\label{master_table_xyper_cont}
\centerline{
\tiny
\begin{tabular}{lllrrl}
\hline
\hline
Line & JD & Event & $v$ (\kms) & \deltav\ (\kms) & $R_{\rm max}$ \\
\hline
\ion{Ca}{ii} K       & 1113.5483 & 7 & -18 & 107 & 0.71 \\
\ion{Ca}{ii} H       & 1113.5483 & 7 & -22 &  83 & 0.81 \\
\ion{Na}{i} D2       & 1113.5483 & 7 &  -4 &  64 & 0.37 \\
\ion{Na}{i} D1       & 1113.5483 & 7 &  -3 &  56 & 0.22 \\
\ion{Fe}{ii} 4924\AA & 1113.5483 & 7 & -10 &  92 & 0.47 \\
\ion{Fe}{ii} 5018\AA & 1113.5483 & 7 & -12 & 101 & 0.42 \\
\ion{Fe}{ii} 5169\AA & 1113.5483 & 7 & -11 &  90 & 0.60 \\
\ion{Ti}{ii} 4444\AA & 1113.5483 & 7 &  -1 & 106 & 0.12 \\
\ion{Ti}{ii} 4572\AA & 1113.5483 & 7 &  -2 & 106 & 0.13 \\
\ion{Ca}{ii} K       & 1113.5483 & 0 &   1 &  17 & 0.27 \\
\ion{Ca}{ii} H       & 1113.5483 & 0 &   1 &  14 & 0.33 \\
\ion{Na}{i} D2       & 1113.5483 & 0 &   1 &  20 & 0.69 \\
\ion{Na}{i} D1       & 1113.5483 & 0 &   1 &  18 & 0.85 \\
\ion{Fe}{ii} 5018\AA & 1113.5483 & 0 & -14 &  27 & 0.17 \\
H$\beta$             & 1207.3786 & 8 & 141 & 120 & 0.16 \\
H$\gamma$            & 1207.3786 & 8 & 121 & 115 & 0.18 \\
H$\delta$            & 1207.3786 & 8 & 118 & 117 & 0.19 \\
H$\zeta$             & 1207.3786 & 8 & 108 & 129 & 0.17 \\
\ion{Ca}{ii} K       & 1207.3786 & 8 & 151 &  58 & 0.06 \\
\ion{Na}{i} D2       & 1207.3786 & 8 &  43 & 117 & 0.16 \\
\ion{Na}{i} D1       & 1207.3786 & 8 &  59 & 159 & 0.10 \\
\ion{Fe}{ii} 4924\AA & 1207.3786 & 8 &  87 & 149 & 0.17 \\
\ion{Fe}{ii} 5018\AA & 1207.3786 & 8 &  99 & 123 & 0.16 \\
\ion{Fe}{ii} 5169\AA & 1207.3786 & 8 &  54 & 196 & 0.27 \\
\ion{Fe}{i} 4046\AA  & 1207.3786 & 8 &  61 & 188 & 0.04 \\
\ion{Ti}{ii} 4444\AA & 1207.3786 & 8 &  46 & 205 & 0.10 \\
\ion{Ti}{ii} 4572\AA & 1207.3786 & 8 &  83 & 144 & 0.08 \\
\ion{Sc}{ii} 4247\AA & 1207.3786 & 8 &  61 & 144 & 0.10 \\
\ion{Ca}{i} 4227\AA  & 1207.3786 & 8 &  40 & 168 & 0.05 \\
H$\beta$             & 1207.3786 & 9 & -13 & 128 & 0.90 \\
H$\gamma$            & 1207.3786 & 9 &  -9 & 132 & 0.84 \\
H$\delta$            & 1207.3786 & 9 &  -7 & 133 & 0.77 \\
H$\epsilon$          & 1207.3786 & 9 &   2 &  72 & 0.53 \\
H$\zeta$             & 1207.3786 & 9 &  -6 & 127 & 0.56 \\
\ion{Ca}{ii} K       & 1207.3786 & 9 & -18 &  92 & 0.71 \\
\ion{Ca}{ii} H       & 1207.3786 & 9 & -16 &  73 & 0.78 \\
\ion{Na}{i} D2       & 1207.3786 & 9 & -23 &  40 & 0.06 \\
\ion{Fe}{ii} 4924\AA & 1207.3786 & 9 & -13 & 125 & 0.20 \\
\ion{Fe}{ii} 5018\AA & 1207.3786 & 9 &  -4 & 111 & 0.28 \\
\ion{Fe}{ii} 5169\AA & 1207.3786 & 9 & -46 &  47 & 0.10 \\
\ion{Ti}{ii} 4444\AA & 1207.3786 & 9 & -21 &  44 & 0.01 \\
\ion{Ti}{ii} 4572\AA & 1207.3786 & 9 & -25 & 113 & 0.07 \\
\ion{Ca}{ii} K       & 1207.3786 & 0 &   0 &  24 & 0.30 \\
\ion{Ca}{ii} H       & 1207.3786 & 0 &   1 &  22 & 0.44 \\
\hline
\end{tabular}
\hspace{1.5cm}
\begin{tabular}{lllrrl}
\hline
\hline
Line & JD & Event & $v$ (\kms) & \deltav\ (\kms) & $R_{\rm max}$ \\
\hline
\ion{Na}{i} D2       & 1207.3786 & 0 &  -0 &  20 & 0.93 \\
\ion{Na}{i} D1       & 1207.3786 & 0 &   0 &  18 & 1.00 \\
\ion{Fe}{ii} 4924\AA & 1207.3786 & 0 &  -2 &  26 & 0.17 \\
\ion{Fe}{ii} 5018\AA & 1207.3786 & 0 &  -2 &  26 & 0.19 \\
\ion{Fe}{ii} 5169\AA & 1207.3786 & 0 &  -4 &  37 & 0.33 \\
\ion{Sc}{ii} 4247\AA & 1207.3786 & 0 &   4 &  18 & 0.04 \\
H$\gamma$            & 1209.3844 & 8 & 109 &  54 & 0.09 \\
H$\delta$            & 1209.3844 & 8 & 102 &  76 & 0.12 \\
H$\epsilon$          & 1209.3844 & 8 &  69 & 123 & 0.11 \\
H$\zeta$             & 1209.3844 & 8 &  84 &  96 & 0.11 \\
\ion{Na}{i} D2       & 1209.3844 & 8 &  42 & 104 & 0.15 \\
\ion{Na}{i} D1       & 1209.3844 & 8 &  55 &  81 & 0.12 \\
\ion{Fe}{ii} 4924\AA & 1209.3844 & 8 &  36 & 145 & 0.20 \\
\ion{Fe}{ii} 5018\AA & 1209.3844 & 8 &  57 & 161 & 0.19 \\
\ion{Fe}{ii} 5169\AA & 1209.3844 & 8 &  53 & 152 & 0.17 \\
\ion{Fe}{i} 4046\AA  & 1209.3844 & 8 &  44 &  88 & 0.04 \\
\ion{Ti}{ii} 4444\AA & 1209.3844 & 8 &  54 & 110 & 0.11 \\
\ion{Ti}{ii} 4572\AA & 1209.3844 & 8 &  55 & 100 & 0.09 \\
\ion{Sc}{ii} 4247\AA & 1209.3844 & 8 &  49 & 117 & 0.11 \\
\ion{Ca}{i} 4227\AA  & 1209.3844 & 8 &  37 & 128 & 0.07 \\
H$\beta$             & 1209.3844 & 9 & -24 & 124 & 0.92 \\
H$\gamma$            & 1209.3844 & 9 & -18 & 129 & 0.83 \\
H$\delta$            & 1209.3844 & 9 & -14 & 118 & 0.73 \\
H$\epsilon$          & 1209.3844 & 9 &  -7 &  50 & 0.32 \\
H$\zeta$             & 1209.3844 & 9 & -14 & 104 & 0.49 \\
\ion{Ca}{ii} K       & 1209.3844 & 9 & -33 &  89 & 0.71 \\
\ion{Ca}{ii} H       & 1209.3844 & 9 & -32 &  74 & 0.77 \\
\ion{Na}{i} D2       & 1209.3844 & 9 & -32 &  45 & 0.18 \\
\ion{Na}{i} D1       & 1209.3844 & 9 & -32 &  25 & 0.09 \\
\ion{Fe}{ii} 4924\AA & 1209.3844 & 9 & -18 &  70 & 0.25 \\
\ion{Fe}{ii} 5018\AA & 1209.3844 & 9 & -15 &  73 & 0.34 \\
\ion{Fe}{ii} 5169\AA & 1209.3844 & 9 & -19 &  78 & 0.40 \\
\ion{Ti}{ii} 4444\AA & 1209.3844 & 9 & -24 &  67 & 0.08 \\
\ion{Ti}{ii} 4572\AA & 1209.3844 & 9 & -19 &  64 & 0.09 \\
\ion{Sc}{ii} 4247\AA & 1209.3844 & 9 & -20 &  39 & 0.05 \\
\ion{Ca}{ii} K       & 1209.3844 & 0 &   1 &  28 & 0.44 \\
\ion{Ca}{ii} H       & 1209.3844 & 0 &   2 &  27 & 0.62 \\
\ion{Na}{i} D2       & 1209.3844 & 0 &   1 &  19 & 0.94 \\
\ion{Na}{i} D1       & 1209.3844 & 0 &   1 &  19 & 1.03 \\
\ion{Fe}{ii} 4924\AA & 1209.3844 & 0 &   0 &  17 & 0.14 \\
\ion{Fe}{ii} 5018\AA & 1209.3844 & 0 &   0 &  17 & 0.16 \\
\ion{Fe}{ii} 5169\AA & 1209.3844 & 0 &   0 &  22 & 0.18 \\
\ion{Ti}{ii} 4444\AA & 1209.3844 & 0 &   3 &  12 & 0.02 \\
\ion{Ti}{ii} 4572\AA & 1209.3844 & 0 &  -2 &   7 & 0.01 \\
\hline
\end{tabular}}
\end{table*}


\end{document}